\documentclass[%
 reprint,
 amsmath,amssymb,
 aps,
]{revtex4-2}

\usepackage{graphicx}% Include figure files
\usepackage{dcolumn}% Align table columns on decimal point
\usepackage{bm}% bold math
\usepackage[colorlinks,citecolor=blue]{hyperref}%\usepackage[mathlines]{lineno}% Enable numbering of text and display math
\usepackage{color}

\usepackage[dvipsnames]{xcolor}
\usepackage[colorinlistoftodos, color=green!40, prependcaption]{todonotes}
\usepackage{soul}
\setstcolor{red}
\setul{1ex}{0.3ex}

\newcommand{\gcc}{g$\,$cm$^{-3}$}
\definecolor{darkviolet}{rgb}{0.58, 0.0, 0.83}

\begin{document}

%\preprint{APS/123-QED}

\title{High Pressure Phase Diagram of Beryllium from \emph{Ab Initio} Free Energy Calculations}

\author{Jizhou Wu}
 \email{wjz8597@berkeley.edu}
\affiliation{Department of Earth and Planetary Science, University of California, Berkeley, CA 94720, USA}%
\author{Felipe Gonz\'alez-Cataldo}
\email{f_gonzalez@berkeley.edu}
\affiliation{Department of Earth and Planetary Science, University of California, Berkeley, CA 94720, USA}%
\author{Burkhard Militzer}
\email{militzer@berkeley.edu}
\affiliation{Department of Earth and Planetary Science, University of California, Berkeley, CA 94720, USA}
\affiliation{Department of Astronomy, University of California, Berkeley, CA 94720, USA}

\date{\today}% It is always \today, today,

\begin{abstract}
We use first principles molecular dynamics simulations coupled to the thermodynamic integration method to study the hcp-bcc transition and melting of beryllium up to a pressure of 1600~GPa. We derive the melting line by equating solid and liquid Gibbs free energies, and represent it by a Simon Glatzel fit $T_m= 1564~\text{K} (1 + P/(15.6032  ~\text{GPa}))^{0.383}$, which is in good agreement with previous two-phase simulations below 6000~K. We also derive the hcp-bcc solid-solid phase boundary and show that the quasiharmonic approximation underestimates the stability of the hcp structure, predicting lower transition pressures between hcp and bcc phases.
Our results are consistent with the stability regime predicted by the phonon quasiparticle method.
We also predict that hcp-bcc-liquid triple point is located at 164.7~GPa and 4314~K. In addition, we compute the shock Hugoniot curve, and show that it is in good agreement with experiments, intersecting our derived melting curve at $\sim$235~GPa and 4900~K. Finally, we make predictions for future ramp compression experiments. Starting with an isentropic compression of the liquid, we predict the path to intersect the melting line at low pressure and temperature, then to continue along the melting line over a large temperature interval of 7000~K as the sample remains in the mixed solid-liquid state before it enters the solid phase. 
%that an isentropic compression path that intersects the melting curve at both low and high pressure and temperature in the liquid regime, can reappear in the solid after a gap as large as 7000~K. Therefore, we predict that a large section of the melting curve could be sampled, in principle, by a ramp compression experiment, where solid and liquid Be would coexist as the sample is compressed.
  %in order to guide quasi-isentropic ramp compression experiments, we derive an isentropic profile that shows a large density gap between liquid and solid after intersecting the melting curve.
% 
% including quasiharmonic approximation (QHA), two phase method, heat until melt(HUM), modified embedded atom model (MEAM) and velocity autocorrelation function (VACF). Our calculation suggests that traditional quasiharmonic approximation method tends to underestimate the hcp-bc
%solid phase boundary. Hugoniot curve was derived from ab-initio molecular dynamics and intersected with melting curve at $\sim$230GPa, 4400 Kelvin. Isentropes and entropy difference near the melting line were also derived to offer a viable reference for future shock experiments.
\end{abstract}

\maketitle

%\tableofcontents

\section{Introduction}
Beryllium (Be) is a widely-used material in space science, plasma physics, and nuclear science because of its high stiffness, low opacity, and high thermal conductivity~\cite{Migliori2004b,Wilson1998}. It serves as an ablator material in internal confinement fusion~(ICF) experiments, as it withstands extreme conditions of several megabar and thousands of Kelvin under shock conditions ~\cite{Clark2008,Haan2011,Simakov2014,Kline2016,Clark2018}. This has triggered a number of studies to investigate its phase diagram, equation of state (EOS) and physical properties. Precise knowledge of the beryllium EOS and phase diagram is of vital importance for understanding the dynamical response of ICF capsules after the shock pulse and to control the growth of hydrodynamic instabilities in the ablator~\cite{Benedict2009,McCoy2019a,Peterson2014}.

Over the past decades, several theoretical and experimental studies have been performed in order to understand the phase diagram and EOS of beryllium. %under different pressures. 
Theoretical studies suggest that at 0~K and high pressure, Be transforms from the hcp to the bcc structure~\cite{mcmahan1982alpha, Lam1984, Meyer-Ter-Vehn1988, Palanivel2000,Sinko2005a, Robert2006a, Luo2012, Benedict2009, Robert2010, Xian2019b, Coe2020}. Calculations using the linear-muffin-tin-orbital~(LMTO) method ~\cite{mcmahan1982alpha} as well as \textit{ab initio} pseudopotential simulations ~\cite{Lam1984} have predicted this transition to occur between 100-200~GPa. Meyer \emph{et al.}~\cite{Meyer-Ter-Vehn1988} implemented  the augmented-spherical-wave method (ASW) in combination with a quantum statistical model and found that the bcc structure would become more stable at $\sim$300~GPa. Palanivel \emph{et al.}~\cite{Palanivel2000} used the full-potential  linear muffin-tin orbital (FP-LMTO) method together with the local density approximation (LDA) and found the transition pressure at 180~GPa. Sinko \emph{et al.}~\cite{Sinko2005a} predicted this transition to occur at 270~GPa using FP-LMTO method with generalized gradient approximation (GGA) functional. Kadas \emph{et al.}~\cite{kadas2007structural} used the exact muffin-tin orbitals (EMTO) method to derive a transition pressure of 240~GPa. Recent predictions from first-principles calculations range from 390 to 420~GPa~\cite{Robert2006a,Benedict2009,Robert2010,Luo2012}, while Coe \emph{et al.} found a transition pressure of 325~GPa~\cite{Coe2020} based on a multiphase equation of state (EOS).

From experiments at room temperature, Ming \emph{et al.} reported a phase transition from hcp to distorted-hcp~\cite{Ming1984a} between 8.6 and 14.5~GPa based on x-ray diffraction measurements, while Vijayakumar \emph{et al.}~\cite{Vijayakumar1984b} 
claimed the existence of a new orthorhombic phase from their measurements of electrical resistivity. However, more recent experiments that employed either x-ray diffraction~\cite{Nakano2002b,Evans2005a,Velisavljevic2002a,Lazicki2012b} or Raman spectroscopy methods~\cite{Olijnyk2000,Evans2005a} confirmed that the hcp phase is stable up to 200~GPa. 

% At ambient pressure, phase transition from hcp to $\beta$-bcc~\cite{Martin1959} or distorted hcp~\cite{Ming1984a, Vijayakumar1984b} structure has been proposed by several studies. When heated up to around 1500 kelvin, it will be subject to a phase transition before melting at 1560 kelvin.
% However recent measurements by either diamond anvil cell (DAC)~\cite{Lazicki2012b} or shock wave experiments~\cite{McCoy2019a} presented no direct evidence for hcp-bcc phase transition,owing to uncertainty of measurement and tiny Gibbs free energy difference near phase boundary.
% 
% It has been shown that dislocations play an important role in the hcp-bcc transition~\cite{Riffet2020}. 
% 

The phase boundary between hcp and bcc Be at 
%high pressure and temperature 
higher temperatures is also a matter of debate. Calculations predict a negative Clapeyron slope along the phase boundary, meaning that the transition temperature decreases with increasing pressure ~\cite{Benedict2009,Robert2010,Luo2012,Xian2019b}.
The quasiharmonic approximation (QHA), a standard method to calculate free energies at high temperature, has been used to study the vibrational properties of Be at high temperature, but this approach does not consider the anharmonic effects~\cite{Robert2010,Luo2012,Xian2019b}. 
Under the QHA, the free energy of solids at high temperatures is obtained from phonon frequencies. Robert \emph{et al.}~\cite{Robert2006a} predicted a solid phase boundary with lower transition temperatures using QHA, with the hcp-bcc-liquid triple point located at 85~GPa and 3400~K. Another work, also based on the QHA, by Luo \emph{et al.}~\cite{Luo2012} reported similar results using the LDA functional. A recent work by Xian \emph{et al.}~\cite{Xian2019b} used a different method, based on phonon quasiparticles, to calculate free energies and found that this phase boundary shifts towards  higher temperatures compared to QHA estimations made by Robert \emph{et al.}, promoting the triple point to 165~GPa and 4200~K. Benedict \emph{et al.}~\cite{Benedict2009} calculated the free energy using QHA and a global EOS model, which led to much higher transition pressures. Regarding experiments at high temperature, Laziki \emph{et al.}~\cite{Lazicki2012b} performed x-ray diffraction measurements in a laser-heated diamond anvil cell to study the hcp phase of beryllium up to 205~GPa and 4000 K, and found no evidence of a bcc phase. The experimental shock Hugoniot curve measured by McCoy \emph{et al.} did not show signatures of a bcc phase either before the onset of melting.

Various simulation methods have been applied to study the melting line of beryllium at high pressure and temperature. Both the heat-until-it-melts~\cite{Robert2010} and the two-phase methods~\cite{Benedict2009} have been used to predict the melting temperature of Be at high pressure, while the Modified Embedded Atom Model (MEAM) has been implemented to explore large-scale phenomena of melting under both hydrostatic and shock compression conditions~\cite{Dremov2015}. Although it is often regarded as an upper limit for the melting temperature, the heat-until-it-melts simulations gave results consistent with two-phase simulations~\cite{Benedict2009,Robert2010}. Experimental data under these extreme conditions remains scarce, making it difficult to verify predictions from the various simulation methods.

Due to the high dispersion in theoretical predictions and scarce experimental data, the intersection between the shock Hugoniot and the melting curve is not well constrained yet. Knudson \emph{et al.}~\cite{knudson2012megaamps} compressed beryllium in the Z-machine, measuring the sound speed along the shock Hugoniot curve from shock waves induced by a magnetically-launched flyer plate. They showed that the shock Hugoniot curve first crosses the hcp-bcc transition line at $\sim$175~GPa, and then intersects the melting line at $\sim$205~GPa. McCoy~\emph{et al.}~\cite{McCoy2019a} performed similar measurements, in which they also identified the onset of melt along the Hugoniot at $\sim$205~GPa, but found no conclusive evidence of bcc phase prior to melting. In this case, the experimental setup was not able to resolve the hcp-bcc solid-solid phase transition, due to the similarity of the sound velocities between the two phases. A recent theoretical work by Coe \emph{et al.}~\cite{Coe2020} found a phase boundary between hcp and bcc phases with lower transition pressures, leading to a  Hugoniot curve intersecting the hcp-bcc phase boundary at 150~GPa and the melting line at 205~GPa. They noticed that the phase transition at 150~GPa was correlated with a decrease in sound speed. 
 
A small region of stable $\beta$-Be (bcc) on the phase diagram, slightly below melting line at low pressure, has been proposed by some authors in previous papers~\cite{Martin1959, Robert2010, Lu2017b}. It has been suggested that prior to melting under ambient pressure, beryllium transforms from hcp to bcc phase at around 1530~K, accompanied by a volume reduction of 6\%~\cite{Martin1959,Pistorius1976,francois1965conference,abey1984pressure}. The slope of this hcp-bcc solid phase boundary at ambient pressure has been reported to be either negative~\cite{Pistorius1976,francois1965conference} or positive~\cite{abey1984pressure} in different studies. Robert \emph{et al.}~\cite{Robert2010,Robert2006a} addressed the existence of this small bcc region below the melting line by monitoring the change of phonon frequencies of the $T_1$ mode at the $N$ point with temperature, while Lu \emph{et al.}~\cite{Lu2017b}, using phonon quasiparticles to describe the anharmonic effects, predicted the boundary of this region to have a positive Clapeyron slope of 41 $\pm$ 4~K/GPa and to disappear at 11~GPa. By contrast, recent x-ray diffraction experiments on diamond anvil cell have not found any evidence for this small bcc region~\cite{Lazicki2012b}.
All these discrepancies motivate further work on the phase diagram of Be, where a proper treatment of the anharmonic effects may be fundamental in order to accurately determine the nature of the hcp-bcc transition at high pressures, as well as the melting curve.

% In this study, we aim to include the whole anharmonic effect using thermodynamic integration(TDI) method and compare our phase diagram with previous research.
In this work, we used the thermodynamic integration technique~\cite{Frenkel1984, Polson2000a} to investigate the phase diagram of beryllium, obtaining the free energy of the hcp, bcc, and liquid phases from first-principles molecular dynamic simulations. 
The thermodynamic integration technique captures the full anharmonicity of the crystal, making this study the first attempt to calculate free energies of beryllium without relying on the quasiharmonic approximations or its extensions. 
%Still, in our implementation, the nuclei are classical. At temperatures below XXX K, nuclei quantum effects are found to be important 
%\burkhard{(compare predictions for hcp-bcc transition from classical and quantum QHA)}. 
%
We compare our resulting solid-solid phase boundary with a recent study based on the phonon quasiparticle method~\cite{Xian2019b} and with other works based on the quasiharmonic approach, demonstrating that the QHA tends to underestimate the stability of hcp phase, lowering the hcp-bcc transition pressure as well as the hcp-bcc-liquid triple point. We also derive the melting line for pressures up to 1600~GPa, where we found a melting temperature of 10000~K, as well as the shock Hugoniot curve, which is found to be in good agreement with shock wave experiments.

Our Hugoniot curve intersects the melting line at 235~GPa and 4900~K, consistent with previous theoretical works of dynamical loading by nonequilibrium molecular dynamics (NEMD), where amorphous~\cite{Thompson2012} or recrystallized structures~\cite{Dremov2015} form well below the equilibrium melting curve. These disordered structures could possibly explain the discrepancy of onset pressure of melt along Hugoniot in shock experiments ($\sim$205~GPa)~\cite{knudson2012megaamps,McCoy2019a}. In addition, we derived an isentrope that intersects the melting line at low pressures, and find that beryllium compressed along this thermodynamic path would span over a large section of melting line, a temperature interval as large as 7000~K. We suggest that the melting curve of Be could be measured, in principle, by a single quasi-isentropic ramp compression experiment, where the solid and liquid phases would coexist as the sample is compressed.

\section{Method}
\label{II}
\subsection{\textit{Ab Initio} Molecular Dynamics}
We performed density functional molecular dynamics (DFT-MD) simulations using the Vienna Ab initio Simulation Package (VASP)~\cite{Kresse1996f}
with projector augmented-wave (PAW)~\cite{P.E.Blochl1994c,Kresse1999b,Kresse1996f} method and a canonical ensemble regulated with a Nos\'e-Hoover thermostat~\cite{Nose1984a,Nose1991}.
To describe the exchange-correlation effects, we used the Perdew–Burke–Ernzerhof (PBE)  functional with generalized gradient approximation~\cite{Perdew1996}. Electronic wave functions are expanded in a plane-wave basis with an energy cut-off as high as 1,000 eV.
The molecular dynamics simulations were performed in 128-~($4\times4\times4$) and 144-atoms ($4\times3\times3$) orthorhombic supercells for bcc and hcp phases, respectively.  Liquid simulations were done in cubic cells with 128 atoms. We considered 400 bands to account for partial electronic  occupations. We chose a time step between 0.7 and 1.0~fs  and total simulation times of at least 2~ps to average the thermodynamic quantities. The error bars were derived from blocking averaging method~\cite{allen-tildesley-87,flyvbjerg1989error}. We use a Monkhorst-Pack grid~\cite{Monkhorst1976b} of $2\times2\times2$ $k$-points to sample the Brillouin zone in our \textit{ab initio} MD simulations.
%\burkhard{Mention k-point correction here!}

\subsection{Thermodynamic Integration}
To determine phase diagram, we calculate the Gibbs free energy for each phase. We used a two-step coupling constant integration (CCI) technique to compute the Helmoholtz free energy~\cite{DeWijs1998,Gonzalez-Cataldo2014}. The full energetics of a solid phase is then described as
\begin{equation} \label{eq:1}
F_{\rm DFT} = F_{\rm Ein} + \Delta{F}_{\rm{Ein} \to \rm{cl}} +  \Delta{F}_{\rm{cl} \to \rm{ DFT}}
\end{equation}
where $F_{\rm Ein}$ is Helmholtz free energy of an Einstein crystal with same density.
 This technique allows to obtain the Helmholtz free energy difference between a DFT system and a reference system for which the free energy is known. We chose the Einstein crystal, where all atoms vibrate with the same harmonic frequency around their lattice sites, as our reference system for the solid. A gas of non-interacting particles was chosen as the reference system when we calculate the Gibbs free energy of the liquid. We performed the calculation of Helmholtz free energy difference between the Einstein crystal and the DFT system in two steps, each involving a TDI integral  
\begin{equation} \label{eq:abc}
\Delta{F}_{a \to b} = \int_{0}^{1}\langle U_{\rm b}(  \textbf{\textit{$r_i$}} )-U_{\rm a}( \textbf{\textit{$r_i$}}) \rangle_{\lambda} \,d\lambda,
\end{equation}
where the angle brackets $\langle...\rangle _{\lambda}$ represents the ensemble average generated in simulations with the hybrid potential $U_{\lambda} = \lambda U_{\rm b}+(1-\lambda)U_{\rm a}$ at constant volume and temperature~\cite{DeWijs1998}. 
%When it comes to the integral from classical to DFT system $\Delta F_{cl \to DFT}$, we constructed spring constant and pair potential of the classical system to match the forces in DFT trajectory
The classical system is governed by a combination of harmonic and pair forces. Both are adjusted to match the forces of a DFT trajectory~\cite{Izvekov2004a,PhysRevLett.104.121101}. After we find the average force between each pair of Be atoms in bins of radial separation, we fit a pair potential using a cubic spline function. Five evenly spread values of $\lambda$ $( 0, 0.25, 0.5, 0.75, 1.0)$ were chosen in order to resolve the integral from DFT to the classical system to complete this thermodynamic step. To compute the Helmholtz free energy difference between the system governed by classical pair forces and the Einstein crystal, namely $\Delta F_{\rm Ein \to cl}$, we performed a thermodynamic integration involving multiple classical Monte Carlo simulations to sample a large number of values of $\lambda$. The Gibbs free energy, $G_{\rm DFT}=F_{\rm DFT}+PV$, is then obtained by adding the pressure term, $PV$.

Frenkel and Ladd~\cite{Frenkel1984} introduced a correction to the free energy of an Einstein crystal to account for the missing degrees of freedom in a solid with a fixed center of mass~\cite{Sun2018,Polson2000a}. However, Navascu\'es \emph{et al.}~\cite{Navascues2010} showed that the magnitude of the actual correction should be much smaller. A recent study of the phase diagram of MgO~\cite{Soubiran2020} showed that the Frenkel correction introduced a significant finite-size error that affected the predicted B1-B2 phase boundary if the simulations of B1 and B2 phases were performed with a small number of particles. Without the Frenkel correction, the results of small and large cells were much more consistent. In this study, we also obtained inconsistent results if we included the Frenkel correction, namely, an overestimation of the stability field of the solid phases that resulted in melting temperatures that are incompatible with previous results~\cite{Robert2006a,Benedict2009}. Therefore, we did not apply this correction to any of our results. We found that our hcp-bcc phase boundary would not be affected by this correction because we used a comparable number of particles for both phases. 

\subsection{K-point correction}\label{sec:FreeEnergyCorr}
We found that a $2\times2\times2$ $k$-point grid in combination with our 128 and 144 atom supercells was not sufficient to obtain converged internal energies. However, using larger supercells or denser $k$-point grids in DFT-MD simulations would be too time consuming. In order to compensate for this drawback, we corrected the unconverged energies using the free energy perturbation (FEP) method, where the internal energy is recalculated for a smaller number of snapshots with a denser $k$-point grid, as explained in Ref.~\cite{allen-tildesley-87}, and used it to correct the free energy by

% \begin{equation} \label{eq:DeltaF}
% F_{\rm DFT}^{\rm high\,k} 
% \approx F_{\rm cl}\\ 
% %&+& 
% \int_{0}^{1} \!\!\!\! d\lambda \, \left[ \, \left\langle  U_{\rm DFT}^{\rm low\,k} - U_{cl} \right\rangle_{\lambda} 
% + 
% \left< U_{\rm DFT}^{\rm high\,k} - U_{\rm DFT}^{\rm low\,k}\right>_{\lambda} \, \right] 
% \end{equation}
\begin{equation} \label{eq:DeltaF}
\frac{F_{\rm DFT}^{\text{high }k} - F_{\rm DFT}^{\text{low }k}}{k_B T}
=- 
\ln\left< \exp\left(-\frac{U_{\rm DFT}^{\text{high }k} - U_{\rm DFT}^{\text{low }k}}{k_{B}T}\right) \right>_{\text{low }k}
\end{equation}
where $F_{\rm DFT}^{\text{high }k}$ stands for Helmholtz free energy derived with the higher number of k-points ($4\times4\times4$) and $F_{\rm DFT}^{\text{low }k}$ for Helmholtz free energy calculated with lower number of $k$-points ($2\times2\times2$). The average $\left\langle \ldots \right\rangle_{\rm low\,k}$ represents an ensemble average that is obtained from the time-averaged MD trajectories that are generated with a smaller number of k-points.
%All MD trajectories with hybrid potential for different $\lambda$ values were generated with the lower number of $k$-points ($2\times2\times2$). 
We took one snapshot every 500 steps for each DFT-MD simulation and rederived the internal energies by performing self-consistent DFT calculations using $4\times4\times4$ (which has been tested to be converged for both solid phases) and $2\times2\times2$ Monkhorst-Pack $k$-point grids. $U_{\rm DFT}^{\text{high }k}$ and $U_{\rm DFT}^{\text{low }k}$ denote the internal energy of those configurations computed with high and low number of $k$-points, respectively. The energy difference is almost the same for all the snapshots, so the free energy correction is close to the arithmetic average of the energy differences. K-point correction has also been implemented in previous theoretical works of up-sampled thermodynamic integration~\cite{Grabowski2009a,Duff2015c}.
%  \textbf{$\leftarrow$WHAT DOES THIS MEAN? DO THEY MENTION A TYPE OF CORRECTION AND WE ARE DOING THE SAME ONE? WHAT IS THE PURPOSE OF MENTIONING IT?} 
We compared both methods and got similar free energy corrections.
%The first thermodynamic average $\left< \ldots \right>_\lambda$ is evaluated for every configuration of these trajectories because it measures the difference between the DFT and classical ensembles (Eq.~\ref{eq:abc}) and its variance is comparatively high. Much smaller is the variance of second thermodynamic average that represents the $k$-point correction to the internal energy and represents the difference between DFT systems with a high and a low number of $k$ points. Therefore, we only evaluate it for small number of configuration. {\bf We found 5 to be a sufficient number.} {\it NEEDS DISCUSSION: We also found this ensemble average is almost independent of the parameter $\lambda$ and may be expressed as
% \begin{equation} \label{eq:DeltaEf}
% \Delta U \approx \frac{1}{n}\sum_i\Delta{U}_i
% \end{equation}
%where n (3$\sim$5) is total number of snapshot configurations we chose. 

\subsection{Phonon Free Energies}

In order to derive the transition pressure between hcp and bcc phases at $T=0$~K, we decompose free energy of the solid into three contributions,
\begin{equation} \label{eq:freeenergy}
F(V,T) = E_0(V) +  F_{i}(V,T) + F_e(V,T),
\end{equation}
where $E_0$ is internal energy of the perfect lattice structure and $F_i$, the thermal contribution of the vibrating nuclei, can be expressed as: 
\begin{equation} \label{eq:phonon}
F_{i}(V,T)  = \sum_{ \textbf{\textit{q}}s}\frac{1}{2}\hbar\omega_{ \textbf{\textit{q}}s} + k_BT\ln\sum_{ \textbf{\textit{q}}s}[1-\exp(-\hbar\omega_{ \textbf{\textit{q}}s} / k_BT  )],
\end{equation}
where $\omega_{ \textbf{\textit{q}}s}$ corresponds to the phonon eigenfrequency with vector  \textbf{\textit{q}} of branch $s$ in the Brillouin zone. The electronic contribution to the free energy, $F_e(V,T)$, is negligible compared  to the other two terms~\cite{Benedict2009,Luo2012,Xian2019b}, so we do not consider it for this calculation. At $T=0$~K, the second term of Eq.~\eqref{eq:phonon} vanishes and only first term contributes. In this study, we applied the density functional perturbation theory (DFPT) method~\cite{baroni2001phonons} to investigate the phonon eigenfrequencies of the solid phases and their corresponding zero-point energy. To perform these phonon calculations, we chose a $k$-point grid of $7 \times 7 \times 7$  with an energy cutoff of 1100~eV, such that the precision in the free energy is within 0.5 meV/atom. Phonon eigenfrequencies are derived from the diagonalization of the dynamical matrix, which was obtained in a $31 \times 31 \times 31$ $q$-mesh grid through Fourier interpolation using the \verb+Phonopy+ software~\cite{phonopy}.

% \textbf{Other approaches to obtain the melting curve include the two-phase method~\cite{Benedict2009}, the Z-method~\cite{Gonzalez-Cataldo2016,Gonzalez-Cataldo2018} and using the Lindemann criterion~\cite{Coe2020}. } 

\section{Results and Discussion}
\label{III}

\subsection{Gibbs free energy and phase diagram}

\begin{figure}[hbt]
\centering
\includegraphics[width=8cm]{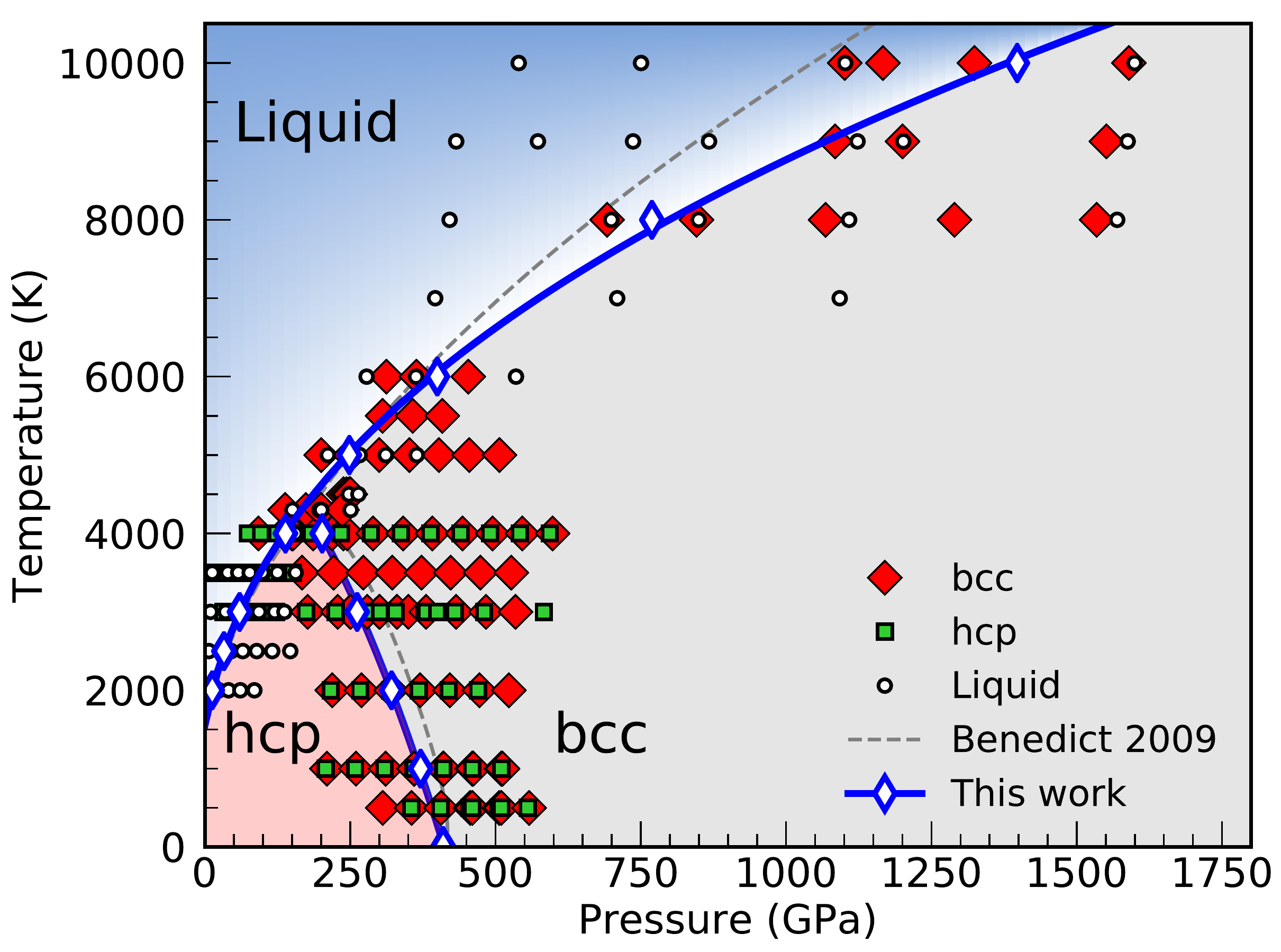}
\includegraphics[width=8cm]{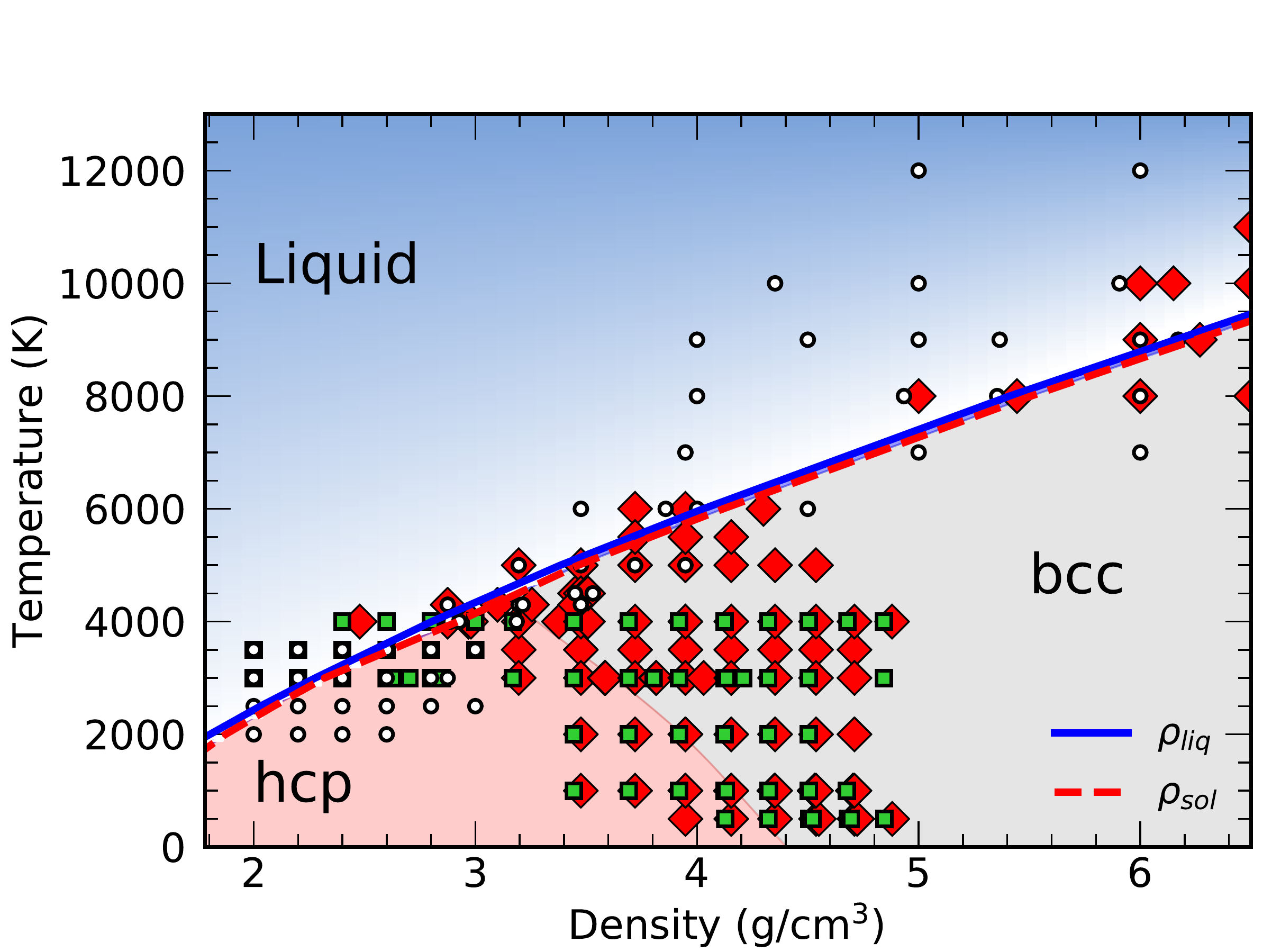}
\caption{Pressure-temperature conditions over which our DFT-MD simulations have been performed. The symbols indicate the different phases explored: liquid (open circles), hcp solid (squares), and bcc solid (diamonds).}
\label{fig:PTconditions}
\end{figure}

\begin{figure}[hbt]
\includegraphics[width=8cm]{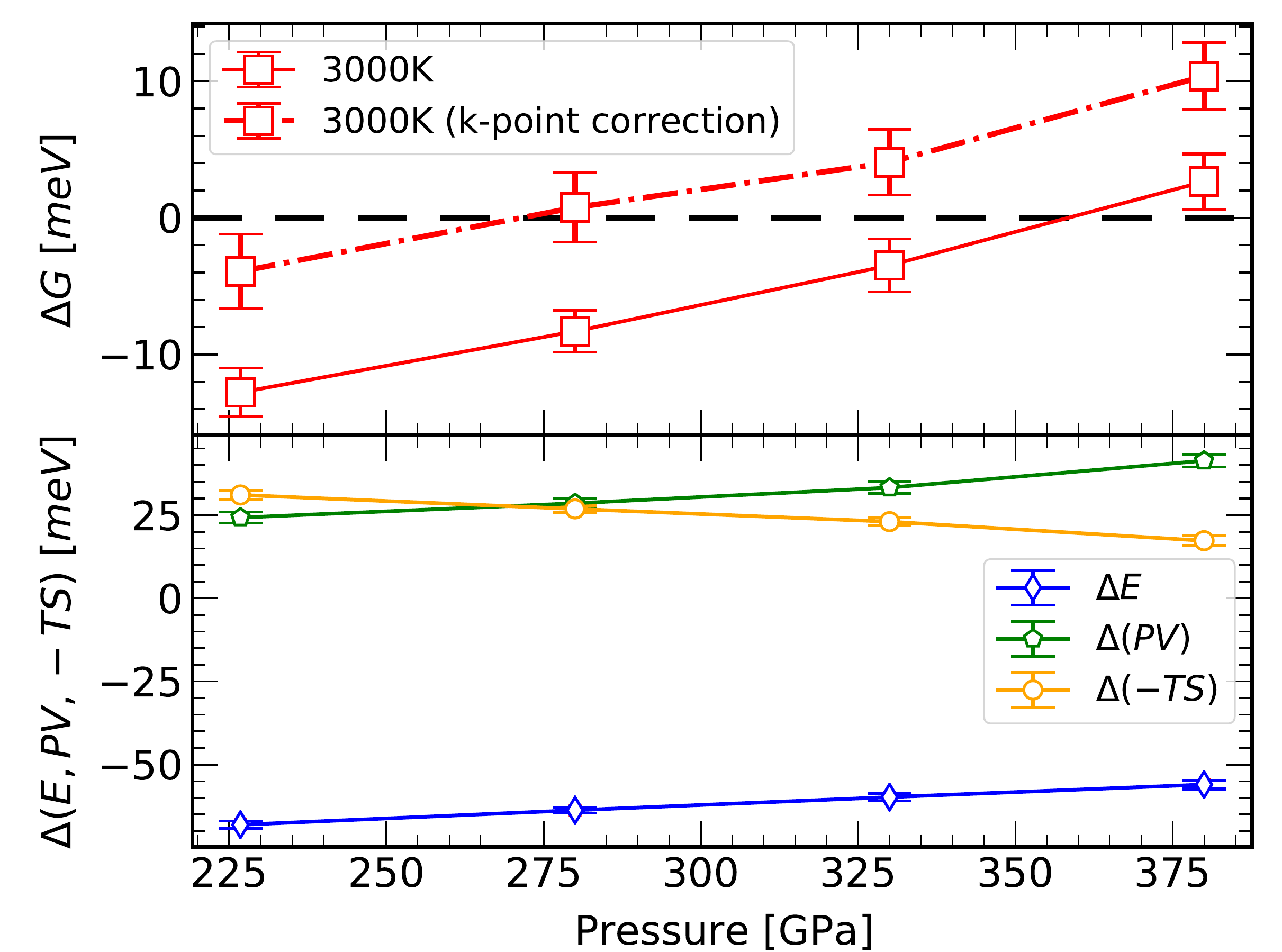}
\caption{Gibbs free energy difference, $\Delta G\equiv G_{\rm hcp} - G_{\rm bcc}$, between hcp and bcc phases along the $T=3000$~K isotherm. The different contributions to $\Delta G=\Delta E+P\Delta V - T\Delta S$ are shown in the lower panel.}
\label{fig:Enthalpydiff}
\end{figure}

We used thermodynamic integration to obtain the Gibbs free energy difference between the different phases of beryllium as a function of pressure for a number of temperatures.
%The pressure at which the Gibbs free energy difference goes through zero makes the transition. 
The pressure at which the Gibbs free energy difference goes to zero marks the phase transition. Fig.~\ref{fig:PTconditions} shows all thermodynamic (density - temperature) conditions where we performed TDI calculations. In Fig.~\ref{fig:Enthalpydiff} we show this difference for a temperature of $T=$ 3000~K as an example. As we observe, the $k$-point correction %described in Sec.~\ref{sec:FreeEnergyCorr}
is necessary, as it shifts the predicted melting pressure to a lower value by more than 50~GPa. The Gibbs free energy difference, $\Delta G\equiv G_{\rm hcp} - G_{\rm bcc}$, increases with pressure and after a critical transition pressure, this difference changes sign and bcc becomes the more stable phase.

The contribution of the different terms $\Delta E$, $P\Delta V$, and $-T\Delta S$ to the Gibbs free energy, as we can see in Fig.~\ref{fig:Enthalpydiff}, shows that the entropic term is comparable to the pressure term, and that the the hcp phase has always lower energy than bcc. 

\begin{figure}[hbt]
\centering
\includegraphics[width=8cm]{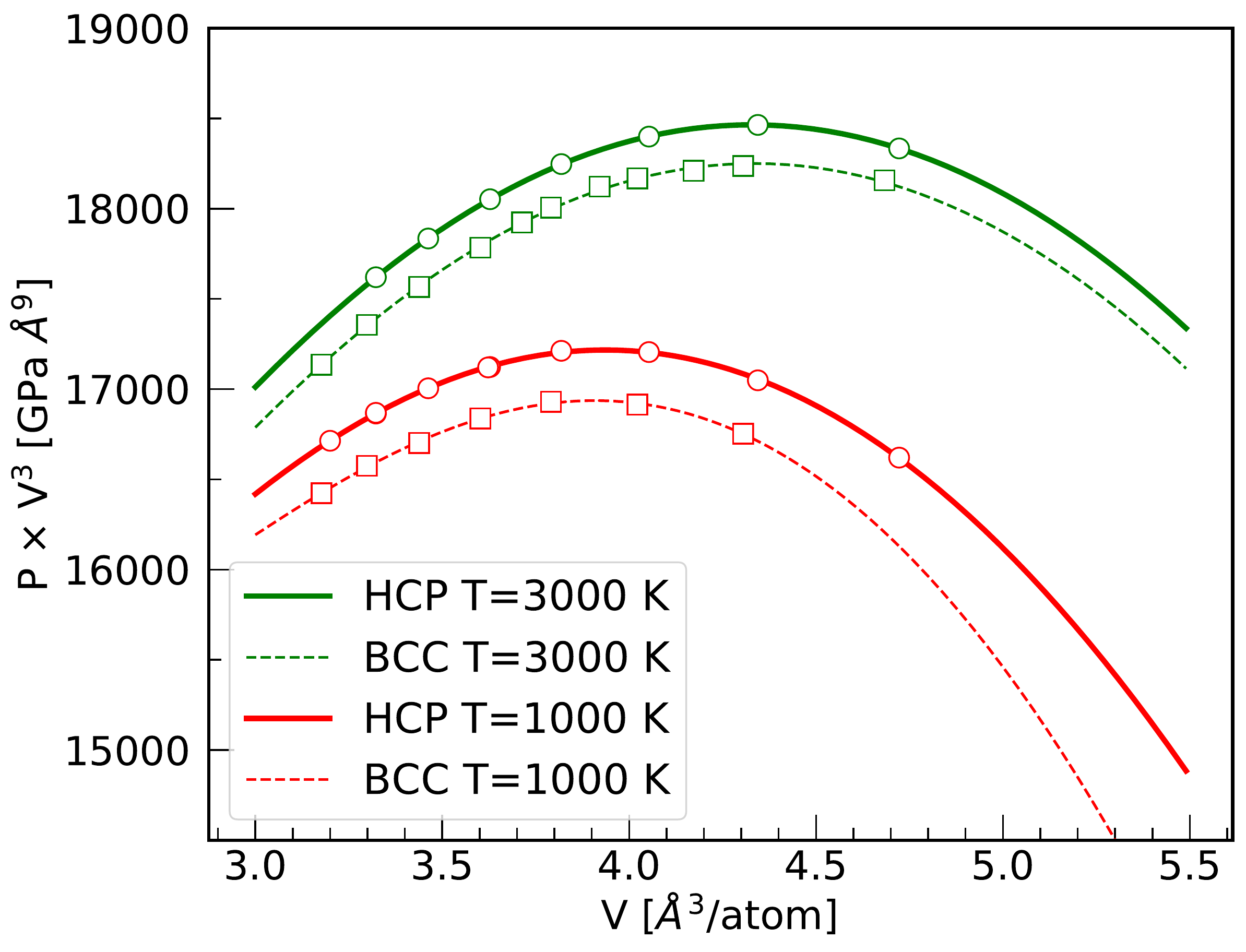}
\caption{Equation of State of hcp and bcc beryllium crystal at $T = 1000$~K and 3000~K. EOS data are fitted to a 4th order Birch Murnaghan Equation of State.}
\label{fig:BMEOS}
\end{figure}

The entropic term contributes with more than 20 meV/atom to the total Gibbs free energy, being crucial in the determination of the hcp-bcc transition pressure. As we can see from the upper panel of Fig.~\ref{fig:Enthalpydiff}, underestimating the Gibbs free energy by 10~meV can make a difference in the transition pressure as large as $\sim$80~GPa. The slope of the Gibbs free energy difference, $(\partial \Delta G/\partial P)_T=\Delta V\approx \Delta G/\Delta P \approx 0.1 \text{ meV/GPa}=0.016$~\AA$^3$ is consistent with the volume difference between hcp and bcc phases that we get from our 3000~K isotherms, $\Delta V=V_{\rm hcp}-V_{\rm bcc}  = 4.080\,{\text \AA^3}-\, 4.064 {\text \AA^3} = 0.016 \,{\text\AA}^3/\text{atom}$ at the transition pressure of $P = 271$~GPa.
%In Fig.~\ref{fig:PTconditions} we show the pressure-temperature conditions that we explored with the DFT-MD simulations. We were able to perform stable simulations of bcc and hcp outside their thermodynamic regime of  stability, as well as under super-cooling liquids and over-heating solids, which allows us to compare their Gibbs free energy at the same pressure and temperature on the phase diagram. 
%
In the lower panel of Fig.~\ref{fig:Enthalpydiff}, we can see that both entropy and pressure terms favor and help stabilize the bcc structure. Since the TDI calculations are performed at constant volume and temperature, a correction must be applied to the Gibbs free energies in order to evaluate both terms of $\Delta G=G_{\rm hcp} - G_{\rm bcc}$ at the same target pressure, $P_T$. This correction is given by $G(P_T,T_0)=G_0+\int_{P_0}^{P_T} V(P)\,dP$, where the integration is performed along the isotherm $T=T_0$ and $G_0=F_0+P_0V_0$ is the Gibbs free energy at the volume $V_0$ chosen for the TDI calculation. We show two of our isotherms $T = 1000$~K and 3000~K in Fig.~\ref{fig:BMEOS}. We chose to plot $PV^3$ as a function of $V$ to enhance the differences in pressure between the two phases which are actually small. At a density $\rho$ = 4 \gcc\, and $T = 1000$~K, the pressures of the hcp and bcc phases are 269 and 264~GPa. When the densities of both phases are compared for a pressure of 300~GPa at 1000 to 3000~K, the bcc phase is found to be 2.0\% and 2.2\%  denser, respectively. The density of the bcc phase is always higher than that of the hcp under the same ($P$,$T$) condition. Overall, we could judge from Fig.~\ref{fig:BMEOS}, bcc phase has a lower $PV$ term in Gibbs free energy at high temperature of thousands of Kelvin.

\begin{figure}[hbt]
\includegraphics[width=8cm]{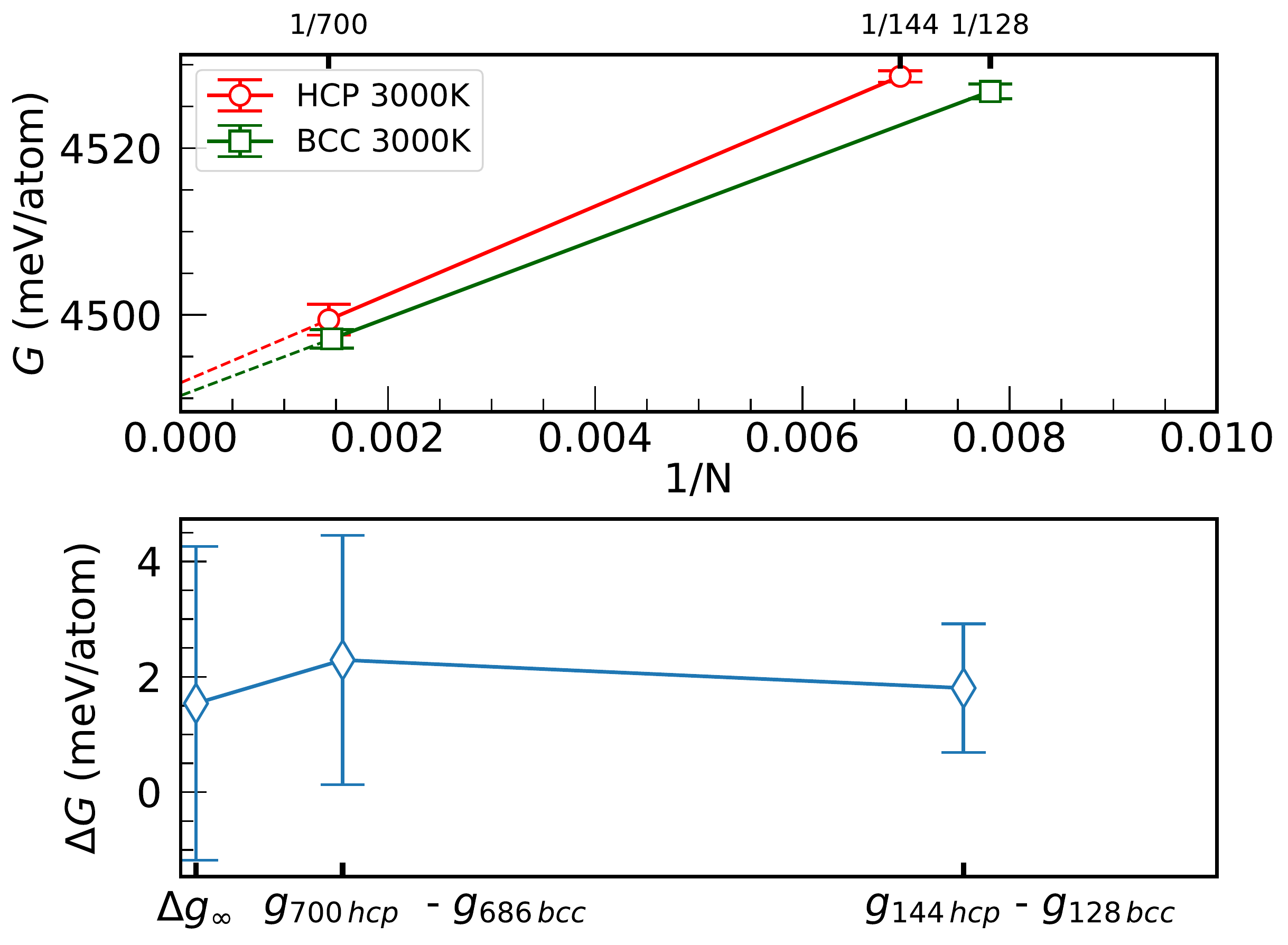}
\caption{Finite size effect on the Gibbs free energy per atom at P = 280~GPa and T = 3000~K.}
\label{fig:Finite_Size_Effect}
\end{figure}

To analyze the finite-size effect of our simulations, we repeated our TDI calculations using larger supercells with 700 and 686 atoms for the hcp and bcc phases, respectively. We used the $\Gamma$ point to sample the Brilloin zone, but applied the $k$-point correction as described in the previous section. We perform our TDI simulation at $P$ = 280 GPa and $T$ = 3000 K as an example. In Fig.~\ref{fig:Finite_Size_Effect}, we shown that the Gibbs free energy decreases with system size. However, when we increased the number of atoms from 128 to 686 in our bcc simulations and from 144 to 700 in our hcp simulations, the Gibbs free energy difference did not change within the error bars. Similarly when we extrapolated our results to infinite size, the resulting Gibbs free energy difference was consistent with those that we originally derived from our simulations with smaller system sizes (see Fig.~\ref{fig:Finite_Size_Effect}). Same conclusion holds for $T$ = 1000~K and $P$ = 410~GPa (see Supplementary Material). Based on these two examples, we conclude that our predictions are sufficiently well converged with respect to system size. 
%  \textbf{THIS IS THE ONLY PARAGRAPH IN THIS SECTION THAT TALKS ABOUT MELTING. MAYBE WE SHOULD MOVE IT TO III.C}
%For a given temperature, we derive the melting pressure by equating the Gibbs free energy of the liquid and solid phases. For temperatures below 4300~K, the solid phase considered is hcp because as we will demonstrate in phase diagram, this phase is more stable than bcc at these conditions. For higher temperatures, bcc becomes more stable, so we compute the Gibbs free energy difference between liquid and bcc phases.
%

%


\begin{figure}[hbt]
\centering
\includegraphics[width=8cm]{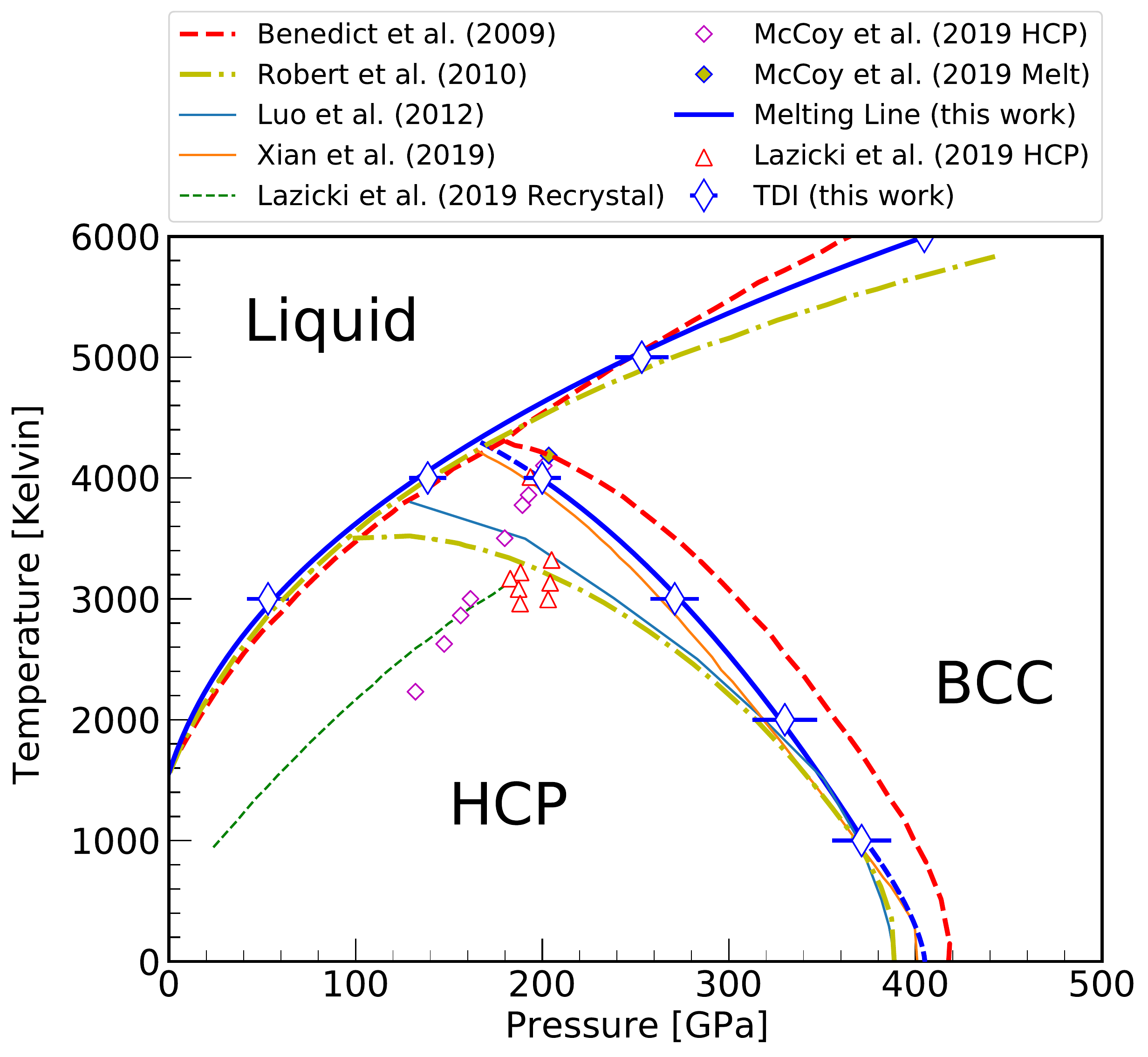}
\caption{Phase boundary of hcp-bcc beryllium, including melting line and solid phase boundary. Blue diamond: solid phase boundary and melting line by TDI(this work); Red upper triangle: HCP phase from diamond anvil cell (DAC) experiments by Lazicki \emph{et al.} ~\cite{Lazicki2012b}; Red dashed curve: phase boundary by two phase method~\cite{Benedict2009}; Yellow dotted dashed line: phase boundary derived by Robert \emph{et al.}~\cite{Robert2010} using QHA and HUM; Blue thin line: solid boundary by Luo \emph{et al.} using QHA (LDA functional)~\cite{Luo2012}; Orange thin line: hcp-bcc solid phase boundary derived by Xian \emph{et al.} using phonon quasiparticle method~\cite{Xian2019b}; Green dashed line: start of recrystallization by Lazicki \emph{et al.} ~\cite{Lazicki2012b}; Magenta open diamonds: hcp beryllium shock wave experiments by McCoy \emph{et al.}~\cite{McCoy2019a}.}
%  \textbf{BENEDICT DASH RED and LAZIKI DASH RED IS CONFUSING!}
\label{fig:PhaseDiagramComplete} 
%  \textbf{FIX LEGEND:  ``HCP,McCoy'' (spaces and consistent with the format LastName (year).}
\end{figure}

In Fig.~\ref{fig:PhaseDiagramComplete}, we show the transition pressures obtained from our TDI calculations and compare our derived phase diagram with previous simulations and experiments. As we can see in the figure, the quasiharmonic approximation (light blue solid line by Luo \emph{et al.}~\cite{Luo2012} and dashed yellow curve by Robert \emph{et al.}~\cite{Robert2010}) underestimates the transition pressure from hcp to bcc beryllium at high temperature.
This phase boundary has also been derived in a recent paper using phonon quasiparticles, fitted from the Fourier transform of velocity autocorrelation function~\cite{Xian2019b}. When anharmonic effects from the phonon quasiparticles are considered, hcp becomes more stable and the transition pressure gets larger compared with QHA. According to the results from their study~\cite{Xian2019b}, the hcp-bcc-liquid triple point is located at 165~GPa and 4200~K, at higher pressure and temperature than those suggested by the QHA method, 85~GPa and 3400~K. Our TDI calculations also point towards similar results, with hcp being more stable compared to QHA results, occupying a larger area of the phase diagram. We will compare our TDI with phonon quasiparticle method and show the differences in the predicted free energies in section~\ref{sec:pdos}. Overall, we will demonstrate that the anharmonic effects of Be, fully captured by TDI, are well-approximated by the phonon quasiparticles below 4000~K. 
The phase diagram of beryllium resembles many features of the phase diagram of MgO, an important material in planetary science, where a B1-B2 solid-solid phase boundary, similar to the hcp-bcc boundary in Be, also exhibits a negative Clapeyron slope~\cite{Boates2013,Bouchet2019,Soubiran2020}. It turns out than anharmonic effects play an important role in promoting B1-B2 transition pressure, especially at high temperature, compared with conventional QHA methods~\cite{Belonoshko2010a,Cebulla2014,Root2015,Miyanishi2015}. To capture anharmonic effects in MgO at high temperature, Boates \emph{et al.}~\cite{Boates2013}  calculated the entropy from the vibrational spectrum derived from the velocity autocorrelation function, 
%Bouchet \emph{et al.}~\cite{Bouchet2019} generated an interatomic force constant (IFCs) matching forces using the TDEP software~\cite{hellman2011lattice}  \textbf{<--NOT CLEAR!} \burkhard{I agree, please reword} 
Bouchet \emph{et al.}~\cite{Bouchet2019} calculated the vibrational free energy using the temperature-dependent effective potential (TDEP) method~\cite{hellman2011lattice}, while Soubiran and Militzer~\cite{Soubiran2020} performed a complete thermodynamic integration. The shape of the B1-B2 solid-solid phase boundary in the MgO phase diagram changes significantly when entropy contributions that go beyond QHA are considered, as anharmonic effects stabilize the B1 phase considerably. Our results for beryllium show that, as it occurs with MgO, the slope of the solid-solid phase boundary is actually steeper than what QHA predicts, which enhances the regime of stability of the low-pressure phase in both cases. This is in agreement with previous diamond anvil cell~\cite{Lazicki2012b} and shock wave ~\cite{McCoy2019a} experiments, where no bcc structure was detected, and all of their measured state points lie within our hcp domain. A fit to our hcp-bcc solid phase boundary leads to a triple point located at 164.7~GPa and 4314~K.

\subsection{Melting curve}
For a given temperature, we derive the melting pressure by equating the Gibbs free energy of the liquid and solid phases. For temperatures below 4300~K, the solid phase considered is hcp because, as we will demonstrate, this phase is more stable than the bcc phase at these conditions. For higher temperatures, the bcc phase becomes more stable, so we compute the Gibbs free energy difference between liquid and bcc phases.
Our TDI calculations also allows us to obtain the Gibbs free energy of the liquid, which we use to obtain the melting curve. At each temperature, we derive the melting pressure by equating the Gibbs free energy of the liquid and solid phases. The melting points obtained from these calculations are shown in Figs.~\ref{fig:PTconditions} and~\ref{fig:PhaseDiagramComplete}.
We fitted our melting curve with the Simon-Glatzel equation~\cite{Simon1953melting},
\begin{equation}\label{eq:SimonFit}
T_m(P)=T_0\,\left(1+\frac{P}{a}\right)^{1/c},
\end{equation}
starting from the experimental value of $T_0=1564$~K~\cite{Martin1959}, and found the parameters $a=15.6032$~GPa and $c=2.6065$.
In Fig.~\ref{fig:MeltingLine}, we compare our melting curve with experiments and other \textit{ab initio} calculations. Predictions from ~\textit{ab intio} simulations using the two-phase~\cite{Benedict2009} and HUM methods~\cite{Robert2010} (pink squares and green triangles in Fig.~\ref{fig:MeltingLine}, respectively) are consistent with our melting points~\cite{Benedict2009,Robert2010}.
%  \textbf{HOW DO THEY KNOW IT IF THEY DID NOT CONSIDER ANHARMONIC EFFECTS?}
% \textbf{Benedict: Heat capacity close to 3k$_B$, limit of Delong Petit; Robert: "pressures obtained from QHA compare well with direct QMD calculations used to determine Tm which seems to indicate that anharmonic effects are not too strong, in agreement with Benedict et al."}
Thus, the predictions from thermodynamic integration and two-phase method agree with each other, at least at low temperatures below 6000~K.

While at low temperatures our melting curve agrees well with the predictions from two-phase simulations, the extrapolated melting line of Benedict~\emph{et al.} (pink dashed line in Fig.~\ref{fig:MeltingLine})~\cite{Benedict2009} results in higher melting temperatures. This difference can be attributed to the fact that only two melting points were reported in their two-phase simulations.
However, the HUM method~\cite{Robert2010}, which is often regarded as the upper limit of melting temperature, also leads to melting temperatures that are consistent with ours at low pressures.
The melting curve fitted to the HUM data from Robert \emph{et al.}~\cite{Robert2010}, which goes below our melting curve in Fig.~\ref{fig:MeltingLine}, shows a large offset with their own data at the highest pressures. For instance, their fitted melting curve shows that the melting temperature at 320~GPa is $\sim$5200~K, which is 800~K below their actual melting data point (bcc phase) and $\sim$300~K below our melting line. 

%the temperature obtained from HUM (green lower triangle in Fig.~\ref{fig:MeltingLine}) is 500~K higher than our melting temperature at 320~GPa, but a little bit higher ($\sim150$~K) than their fitted melting curve. 

A recent study that used both the hysteresis method (HM) as well as thermodynamic integration with a modified embedded atom model (MEAM) parameterization~\cite{Dremov2015} reported a melting line that is several hundred Kelvin above ours. This shows that this empirical potential~\cite{Baskes1987a,Baskes1989a,Baskes1992} cannot fully capture the atomic interactions as well as DFT. Conversely, a recent study based on an EOS model~\cite{Coe2020} proposed a melting line lower than all the reported melting curves so far. They compared their theoretical EOS predictions with DAC experiments~\cite{Lazicki2012b} and showed that discrepancies in the EOS appear along the isochores at high temperature, implying that their melting line should be steeper than what they predict~\cite{Coe2020}. Overall, we obtain a melting curve that is in reasonable agreement with previous predictions from two-phase and HUM simulations, and we extended it to much higher pressures.
% \textbf{COMPARE WITH REF.~\cite{Thompson2012} and the MEAM simulations of~\cite{Karavaev2016} and ~\cite{Dremov2015}.}

% \textbf{ In Ref.~\cite{Dremov2015}, the authors used the Hysteresis Method to determine the melting curve from ab initio simulations, in good agreement with previous works~\cite{Benedict2009,Robert2010}}

% \textbf{COMPARE AND DISCUSS WITH THE WORK OF HANSHAW~\cite{Hanshaw2012}}.

% \textbf{It has been suggested that in the micron-size samples, the phenomenon of cold melting strongly affects the Hugoniot (read more in Ref.~\cite{Dremov2015}) }

% \textbf{In Ref.~\cite{Coe2020}, authors build an EOS model based on free energy calculations and obtain a melting line using the Lindeman criterion, being this work the lowest estimation for the melting temperature among the cited works.}

% \textbf{ 09/25/2020: Calculations by Robert et al. fit a single melting line to all their points, obtained but HUM method, but highest pressure results are above their fitted melting line, closer to the fit done by Benedict and ours. They did not consider the high pressure bcc transition, so that may be the explanation for these outlier points.}

%%%%%%%%%%%%%%%%%%%%%%%%%%%%%%%%%%
%%%%%%%%%%%%%%%%%%%%%%%%%%%%%%%%%%
\subsection{Transition pressure at $T=0$~K}

%\subsection{c/a ratio of hcp beryllium}

\begin{figure}[hbt]
\centering
\includegraphics[width=8cm]{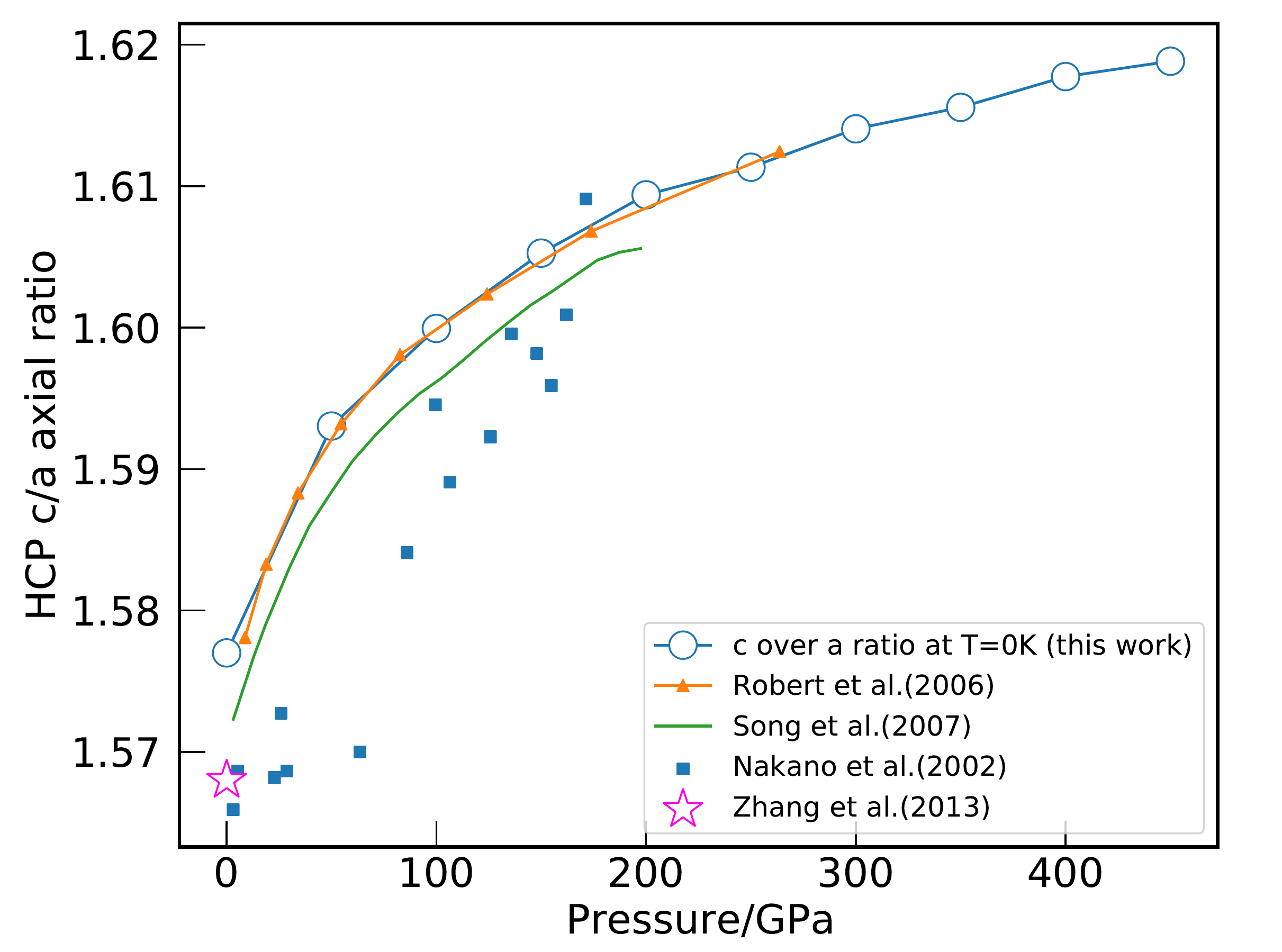}
\caption{Variation of the $c/a$ ratio of hcp phase of beryllium with pressure. Open circle: DFT calculation in this work; Yellow triangle: first principle calculation by Robert \emph{et al}~\cite{Robert2006a}. Green curve: classical analytic mean-field potential method by Song \emph{et al}.~\cite{Song2007}; Blue square: X-ray diffraction DAC experiment by Nakano \emph{et al}.~\cite{Nakano2002b}; Magenta star:experimental value at $P=0$~GPa from Zhang \emph{et al}.~\cite{Zhang2013}.}
\label{fig:c_over_a}
\end{figure}

We derived the $c/a$ ratio of the hcp structure at zero Kelvin as a function of pressure, which is shown in Fig.~\ref{fig:c_over_a}.
As we can observe, as compression increases the $c/a$ ratio rises, converging to the ideal value
of $\sqrt{8/3}\approx 1.633$. At ambient pressure and zero temperature, we obtain a $c/a$ ratio of 1.577, in close agreement with the experimental value 1.568~\cite{Evans2005a,Lazicki2012b,Zhang2013}. The deviation from the ideal value at ambient conditions can be attributed to the large hybridisation of s and p orbitals~\cite{Simak2000,Haussermann2001a}. As pressure increases, the $p_x$, $p_y$ and $p_z$ bands tend to become degenerate, making Be closer to ideal hcp rigid packing at high pressures~\cite{Haussermann2001a,Sinko2005a,Robert2006a}. In our simulations, we assume $c/a$ ratio is constant along each isochore, because the value of this ratio has little impact on the calculated free energy of hcp Be~\cite{Xian2019b}.

\begin{figure}[hbt]
\centering
\includegraphics[width=8cm]{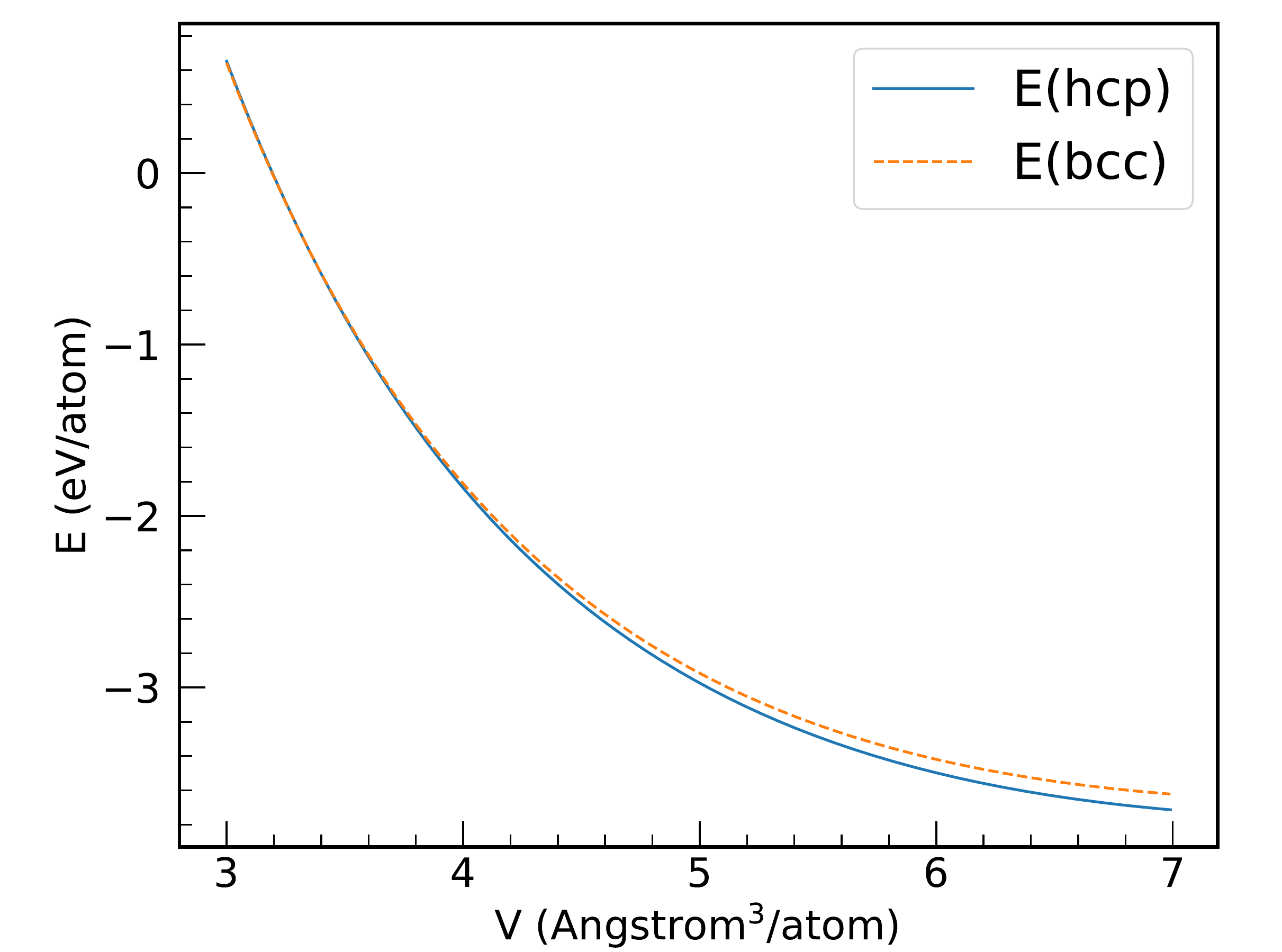}
\caption{Fitted 4th order Birch-Murnaghan equation of state of hcp and bcc beryllium.}
\label{fig:EOS_0K}
\end{figure}

% We derived the equation of state (EOS) from our simulations. Isotherms for the hcp and bcc crystal were fitted to a 4th order Birch Murnaghan EOS at high temperature, which is shown in Fig.~\ref{fig:BMEOS}.
% We observe that bcc is always denser than hcp at all temperatures.

We calculated the energy of the bcc and hcp structures as a function of pressure at $T=0$~K to determine where the phase transition occurs. We used a 
dense $k$-point grid ($43\times43\times43$) to sample the Brillouin zone of the primitive cell and included the zero point energy by performing DFPT phonon calculations.
We fit the cold curve of hcp and bcc beryllium with a 4th order Birch-Murnaghan equation of state, as shown in Fig.~\ref{fig:EOS_0K}. We observe that the energy of hcp phase of beryllium is lower compared to bcc, but the difference decreases as pressure increases. We found that $V_0$ = 7.99 ${\AA}^3$ and the bulk modulus of Be  at $P=0$~GPa in the hcp phase is $B_0$ = 112.96~GPa and $B_0^\prime$ = 3.61, close to previous experiments~\cite{Nakano2002b,Migliori2004b,Evans2005a,Lazicki2012b} and theoretical predictions~\cite{Velisavljevic2002a,Sinko2005a,Robert2006a,Song2007,Benedict2009,Robert2010}. From our fitted EOS for the bcc phase, we find $V_0 = 7.92 $~\AA$^3$, $B_0 = 111.52$~GPa and $B_0^\prime =$ 3.64.
We determined the relative enthalpy between the two phases and find that the transition from hcp to bcc occurs at  405~GPa, consistent with recent theoretical predictions~\cite{Luo2012, Benedict2009,Robert2010, Xian2019b}~  \textbf{(see Fig.}~\ref{fig:PTconditions}).

%\subsection{Hugoniot and Isentrope}
\subsection{Hugoniot calculations and EOS}
\begin{figure}[hbt]
\centering
\includegraphics[width=8cm]{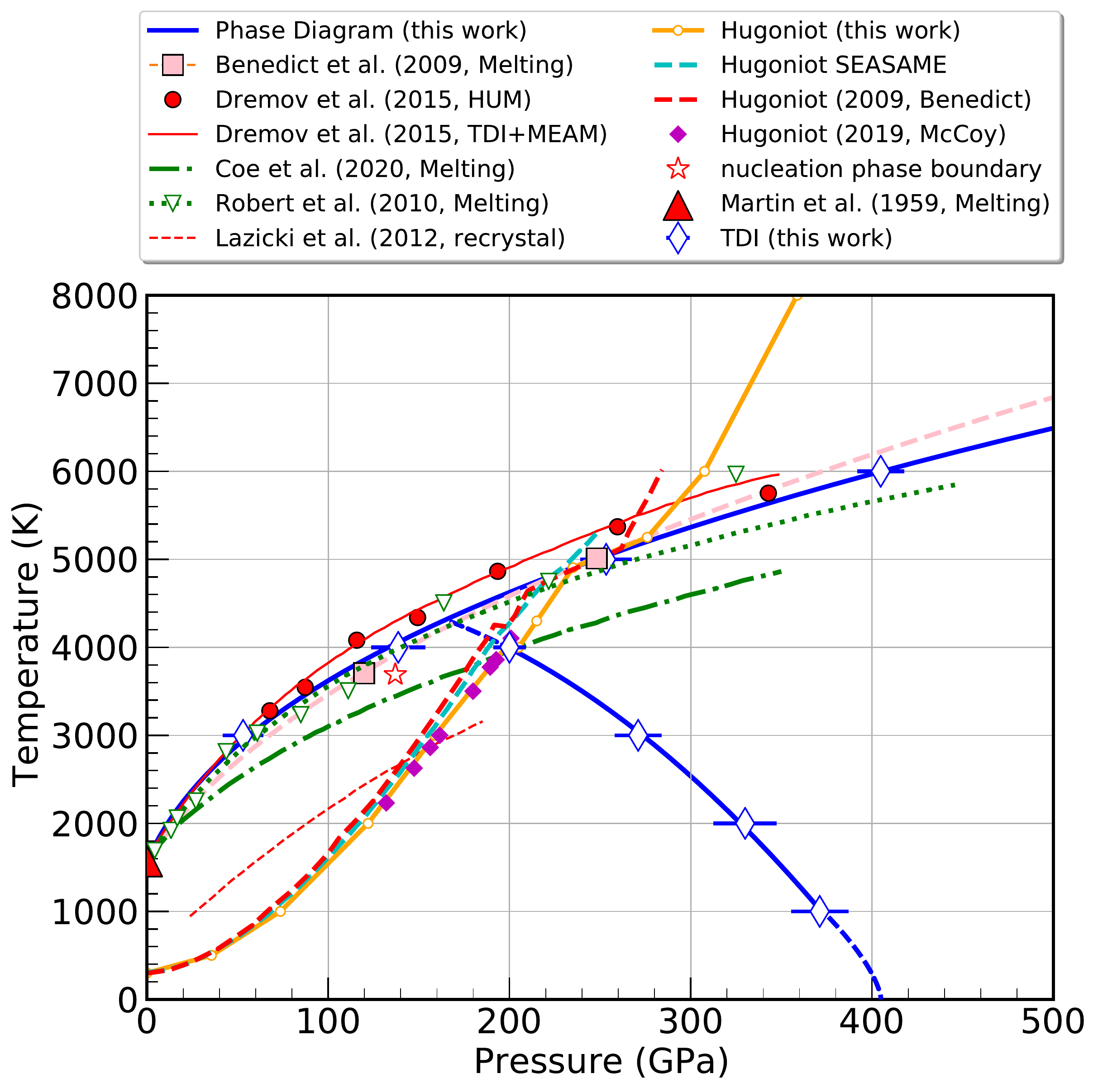}
\caption{Melting line and shock Hugoniot curve of Be. Blue diamond: phase diagram by TDI in this work; Orange thick line: Hugoniot by ab-initio MD in this work; Upper red triangle: Melting temperature at ambient pressure~\cite{Martin1959}; Cyan dotted curve: Hugoniot by SEASAME 2024~\cite{lyon1992sesame}; Dashed red curve: Hugoniot by Benedict \emph{et al.}~\cite{Benedict2009}; Pink squares: melting temperature by Benedict  \emph{et al.} using two-phase method~\cite{Benedict2009}; Green lower triangle: heat-until-it-melts by Robert \emph{et al.}~\cite{Robert2010}; Red circles: melting temperature by Dremov \emph{et al.} using heat-until-it-melts method~\cite{Dremov2015}; Red curve: TDI by Dremov \emph{et al.} using MEAM model~\cite{Dremov2015};  Green dotted line: melting line by Coe \emph{et al.}~\cite{Coe2020}; Open star: point along the nucleation phase boundary derived from classical nucleation theory. %  \textbf{USE TRIANGLES INSTEAD OF SQUARES to make it consistent with Fig. 1 and 3. Use markeredgewidth=2, lw=2.}
}
\label{fig:MeltingLine}
\end{figure}

In Fig.~\ref{fig:MeltingLine}, we show the shock Hugoniot curve of Be that we have obtained by solving the Rankine-Hugoniot condition
\begin{equation}
\label{HugoniotF}
(E-E_0) + \frac{1}{2}(V-V_0)(P+P_0)=0,
\end{equation}
where $E_0$, $V_0$ and $P_0$ are the internal energy, volume and pressure of hcp phase of beryllium at ambient pressure and 300~K.
Our Hugoniot curve is in good agreement with shock wave experiements~\cite{McCoy2019a} and with predictions from other theoretical works~\cite{Benedict2009,Robert2010}. Our Hugoniot curve intersects the hcp-bcc phase boundary at around 200~GPa and 4000~K, showing a very small offset due to the phase transition. The intersection with the melting line occurs at 235~GPa and 4900~K, and the Hugoniot curve reappears in the liquid region around 276~GPa, with an offset of around 40~GPa.
Shock experiments suggest that the onset of melting along the Hugoniot curve occurs at 205~GPa and around 4000~K, based on the criteria that the longitudinal and bulk sound speed are equal~\cite{knudson2012megaamps,McCoy2019a,Coe2020}. This leads to melting temperatures lower than ours, but this criterion may not represent a valid condition for melting at equilibrium.
It has been suggested that this could be attributed to a phenomenon called ``cold melting''~\cite{Thompson2012,Levitas2012,Budzevich2012,He2013,Dremov2015}. In cold melting, disordered structures such as recrystallized grains or amorphous solid form right after the shock front, leaving behind a metastable system instead. Dremov \emph{et al}.~\cite{Dremov2015} considered this effect and corrected the shock Hugoniot curve using large-scale MD simulations, resulting in a intersection with the melting line around 250~GPa and 5000~K~\cite{Dremov2015}, consistent with our simulations. 

Hugoniostat MD simulations by Thompson \emph{et al.}~\cite{Thompson2012} predict that Hugoniot crosses melting line at 230~GPa and 5000~K, consistent with our result.

We also derived an isentrope for beryllium from the entropies that we obtain from our TDI simulations. This is relevant to ramp compression experiments, where the compression is assumed to follow a quasi-isentropic path~\cite{Swift2008,Brown2014a}. If the sample is isentropically compressed from the liquid, it will hit the melting line and remain in a solid/liquid mixture until the pressure is high enough to solidify the sample entirely. If the sample is ramp-compressed further, it will follow a solid isentrope, unless plastic work heating increases the temperature to a significant degree. Here we compute the thermodynamic path of such a ramp compression experiment. As initial conditions, we considered liquid Be at 2000 K and 5~GPa ($\rho=1.66$~\gcc), where we obtained an entropy of $S=7.9\,k_B$/atom from our TDI calculations. 
% \textbf{For the isentrope of the bcc phase of beryllium we start with TDI simulation at 1590~GPa ($\rho=7.00$~\gcc).} 
Then, using the EOS table that we have generated with our simulations, we solve the thermodynamic equation 
\begin{equation} \label{eq:8}
\left(\frac{\partial T}{\partial V}\right)_S=-T\frac{\left(\frac{\partial P}{\partial T}\right)_V}{\left(\frac{\partial E}{\partial T}\right)_V}
\end{equation}
to generate isentrope $S=7.9 \, k_B$/atom for both solid and liquid phases.

\begin{figure}[hbt]
\centering
\includegraphics[width=8cm]{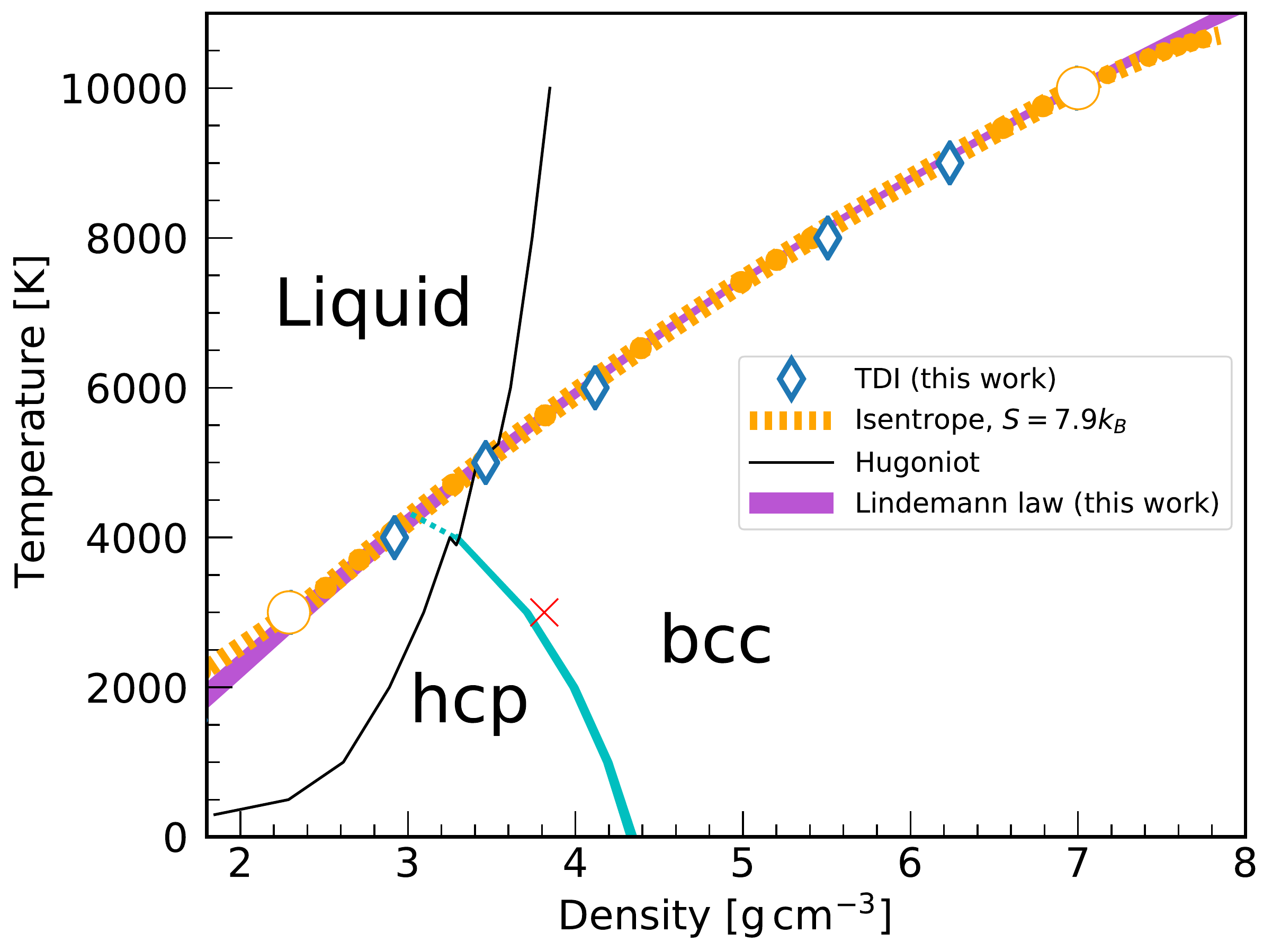}
\caption{ Hugoniot and isotherm of $S = 7.9 \, k_B$/atom of beryllium in density-temperature space. Blue diamonds: melting points obtained from TDI (average density between solid and liquid at the melting temperature); Purple curve: 
melting curve obtained from the Lindemann criterion (see section~\ref{sec:Lindemann}); Orange curve: isentrope S = 7.9 k$_B$/atom of solid and liquid phases by TDI; Cyan curve: hcp-bcc solid phase boundary; Black curve: Shock Hugoniot derived from ab-initio MD; Red cross: density and temperature we investigate using both TDI and phonon quasi particle method.}
\label{fig:isentrope}
\end{figure}

As we can see in Fig.~\ref{fig:isentrope}, this isentrope intersects the melting line at 3000 K and reappears in the solid bcc region at around 10\,000~K ($1590$~GPa, 7.00~\gcc), a temperature gap of 7000~K.
The slope of the melting line is steeper than that of the isentrope, suggesting that isentropic compression should always encounter partial crystallization if pressure is high enough, and that recrystallization of Be during isentropic release in shock decay experiments should not be observable~\cite{Militzer2013d,Davies2020}.

%This means that ramp compression experiments of liquid beryllium would not observe recrystalization of liquid Be unless they are able to reach a compression ratio of 3.7-fold from ambient conditions. This could work as a reference for ramp compression experiments in the future. 

These results imply that isentropic ramp compression experiments should generate a solid-liquid mixture and remain in such a state over a $T$, $\rho$, $P$ interval of approximately 7000~K, 4.7~\gcc\, and 1550~GPa before the mixture freezes  completely at a compression ratio of 3.05-fold from ambient density. The intensity of the X-ray diffraction peaks would surge as the fraction of the solid increases at higher compression. On the other hand, the Debye-Waller effect~\cite{Graf2004,wallace1972thermodynamics} would broaden the peaks. Nevertheless, a long section of the melting line could, in principle, be measured with a ramp compression experiment if accurate temperature measurements become available.
%Thus if we manage to ramp compress heated liquid beryllium we would possibly see a jump of more than 6000 kelvin in Z-machine, which may be tested in future experiment. 

%\subsection{Clapeyron Slope}

\begin{figure}[!hbt]
\centering
\includegraphics[width=8cm]{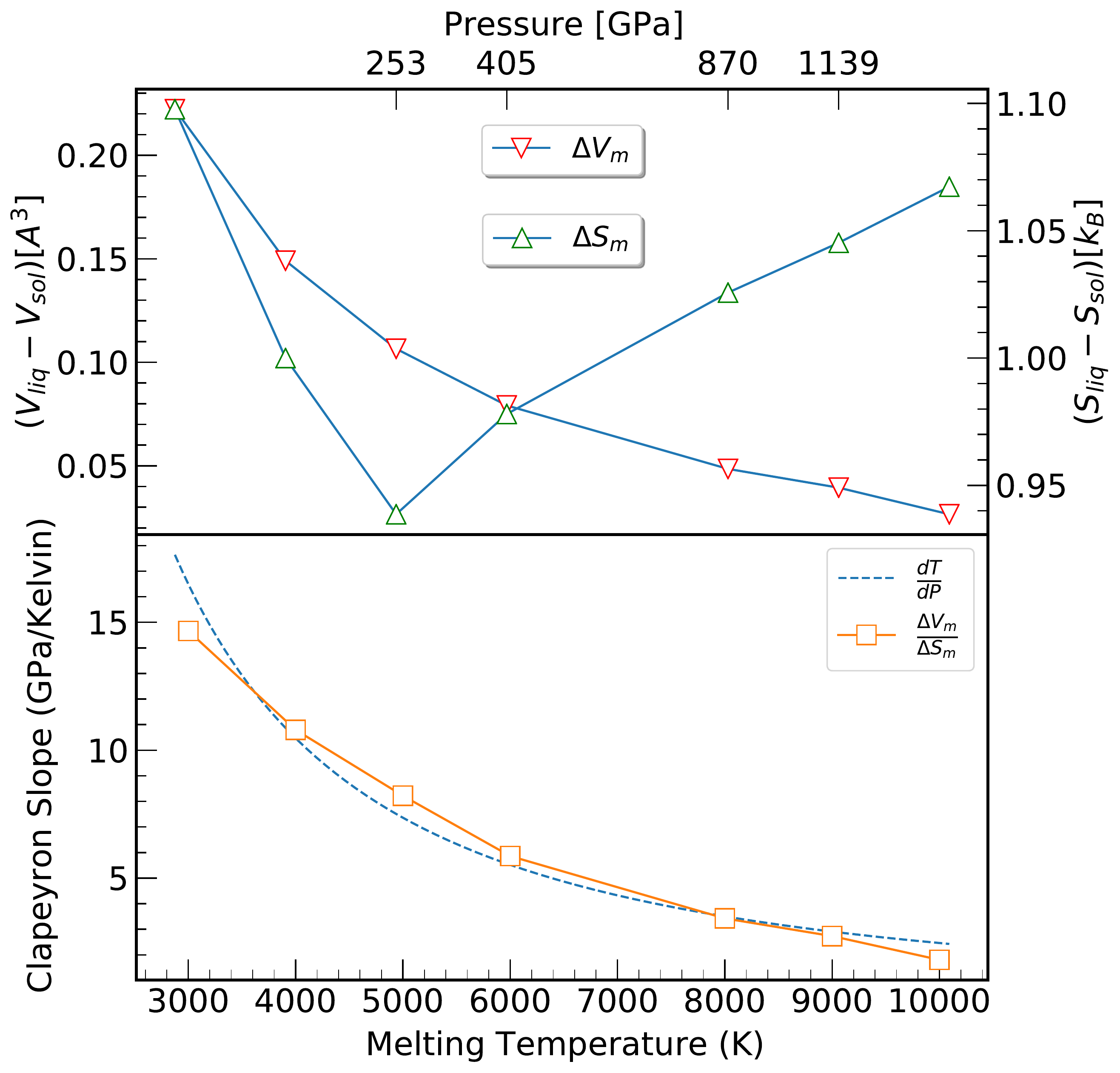}
\caption{Upper panel: Entropy and specific volume difference between Liquid and Solid beryllium along melting line. Lower panel: Comparison between slope of melting line and Clayperon formula.}
\label{fig:DeltaMeltingLine}
\end{figure}
In Fig.~\ref{fig:DeltaMeltingLine}, we show how the volume and entropy differences between the solid and liquid phase change along the melting line and compare their ratio with the slope of our fitted melting curve, as given by the Clausius–Clapeyron relation,
\begin{equation}\label{eq:dTdP}
\left. \frac{dT}{dP} \right|_{\rm m}=\frac{V_{\rm liq} - V_{\rm sol} } {S_{\rm liq} - S_{\rm sol} }\;.
\end{equation}
We obtain consistent results from both approaches, which differ by less than 10\% in the pressure range investigated.

From the entropy of fusion in Fig.~\ref{fig:DeltaMeltingLine}, we can estimate the kinetic (nucleation) effects during the solidification of liquid beryllium at high pressure. It has been reported that a thermodynamically metastable crystal phase may dominate the initial growth of a solidifying cluster in the liquid~\cite{desgranges2007controlling,sadigh2021metastable}. The Gibbs free energy of the solid cluster in the liquid during supercooling is determined not only by the thermodynamic bulk free energy but also by the interfacial Gibbs free energy, $\gamma_I$, which is proportional to the characteristic Turnbull coefficient, $\alpha$, in the Gibbs-Thompson (GT) limit~\cite{turnbull1950formation},
\begin{equation}\label{eq:interfacial free energy}
    \gamma_I = \alpha\frac{\Delta H_m}{v_S^{2/3}} A_I^{\rm sphere},
\end{equation}
where $A_I^{\rm sphere} = (36 \pi)^{1/3}(N_s v_s)^{2/3}$ represents the surface area of the cluster in equilibrium with the surrounding liquid. $v_S$ is the specific volume of the solid. The Turnbull coefficient of hcp, $\alpha_{\rm hcp}$, is assumed to be higher than that of the bcc phase~\cite{sun2004crystal,sadigh2021metastable}. Considering the size limitation of current ab initio MD simulations, as well as the scarcity of data on the Turnbull coefficient of different beryllium phases, it is challenging to calculate the kinetic nucleation boundary with high precision. We estimate $\alpha_{\rm hcp}/\alpha_{\rm bcc}$ = 1.15 based on previous studies that predicted the Turnbull coefficient of the bcc crystals to be approximately 10–20\% lower than that of fcc or hcp crystals~\cite{sadigh2021metastable}. 

% Therefore the total Gibbs free energy of the system could be expressed as
% \begin{equation}\label{eq:Gibbs free energy of the system}
%     G_{SL} = N_S G_S + (N-N_S) G_L + \gamma_I
% \end{equation}
% with $G_S$ and $G_L$ standing for Gibbs free energy of bulk solid and liquid respectively.

% Minimizing the total Gibbs free energy expressed above we have critical size of the cluster 
% \begin{equation}\label{eq:critical cluster size}
%     N_{S}^c(P,T) \approx -\frac{2}{3}\frac{T_m}{\Delta H_m}\frac{\gamma_I}{\Delta T}
% \end{equation}
% where $\Delta H_m$ is enthalpy of fusion and $\Delta T = T_m - T$ is the undercooling temperature. 
The kinetic phase boundary is defined as the state where two phases share the same nucleation rate. In classical nucleation theory (CNT), this rate can be expressed as~\cite{sadigh2021metastable},
\begin{equation}\label{eq:classical nucleation rate}
    J_S(T_c) = \sqrt{\frac{\Delta G_S^{\prime \prime}(T_c)}{2\pi k_B T_c}}[N_S(T_c)]^{2/3}\frac{\tau}{v_L}\exp\left(-\frac{\Delta G_S(T_c)}{k_B T_c}\right),
\end{equation}
where $\tau$ is the rate of attachment to a unit area of the cluster, which we set equal for both phases, $\tau_{\rm hcp} = \tau_{\rm bcc}$. The excess Gibbs free energy, $\Delta G_S$, at the temperature $T_c$ is defined as
\begin{equation}\label{eq:excess Gibbs free energy}
    \Delta G_S(T) = N_S\frac{\Delta H_m}{T_m}(T-T_m) + \alpha (36\pi N_S^2)^{1/3}\Delta H_m,
\end{equation}
where $\Delta G^{\prime \prime}_S$ is the curvature of the excess Gibbs free energy with respect to cluster size. 
Minimizing the Gibbs free energy, we obtain the critical cluster size, 
\begin{equation}\label{eq:critical cluster size}
    N_S = \frac{32\pi}{3}\left(\frac{\alpha T_m}{T_m-T}\right)^3.
\end{equation}
After equating the nucleation rates of both phases, $J_S^{\rm hcp}(T^*) = J_S^{\rm bcc}(T^*)$, we derived one point along the kinetic phase boundary (see open star near the triple point in Fig.~\ref{fig:MeltingLine}). The bcc phase is predicted to exhibit a larger nucleation rate than the thermodynamically preferred hcp phase and may thus dominate the initial nucleation process. The nucleation temperature, $T^*$, may be expressed by 
\begin{equation}\label{eq:nucleation phase boudary}
    \frac{T^* - T_m^{\rm hcp}}{T^* - T_m^{\rm bcc}} = \frac{T_m^{\rm hcp}}{T_m^{\rm bcc}}\sqrt{\left(\frac{\alpha_{\rm hcp}}{\alpha_{\rm bcc}}\right)^3\frac{\Delta H_m^{\rm hcp}}{\Delta H_m^{\rm bcc}}}.
\end{equation}
We derived one point on the hcp-bcc nucleation boundary: $P$= 137~GPa and $\sim$3700~K, just 300~K below the melting curve. Above this boundary, the hcp phase crystallizes more quickly, while when the sample is cooled rapidly to a temperature below this boundary, the bcc polymorph is predicted to form.

%%%%%%%%%%%%%%%%%%%%%%%%%%%%%%%%%%
%%%%%%%%%%%%%%%%%%%%%%%%%%%%%%%%%%
\subsection{Gap along the isentrope derived from the logarithmic phonon moment}\label{sec:Lindemann}

In this section, we provide a simple method that allows us to obtain an approximated expression for the temperature difference between the solid and liquid entropies without relying on expensive TDI calculations. This difference is important to understand the relationship of isentropes and the melting line that is relevant for ramp compression experiments, which are assumed to be quasi-isentropic. Starting with a liquid isentrope, such experiments may intersect and follow the melting line, as the sample remains in a solid-liquid mixed state before it enters the solid phase. 

We start with the assumption that, at low pressures, beryllium has an entropy of fusion of
$\Delta S\equiv S_{\rm liq} - S_{\rm sol} \approx 0.9\,k_B$/atom~\cite{german1996sintering,yefimov2009handbook} close to the ``universal'' entropy of fusion of 0.8 $k_B$ suggested by Wallace~\cite{wallace1972thermodynamics}.

When the isentrope intersects the melting line at a given temperature $T_1$, the liquid has density $\rho_1$ and an entropy $S_{\rm liq}(\rho_1,T_1)$. The thermodynamic path with the same entropy in the solid regime appears at a higher density $\rho_2$ with a temperature $T_2$, and $S_{\rm sol}(\rho_2,T_2)=S_{\rm liq}(\rho_1,T_1)$. Since we assume that the entropy of fusion is known, we can calculate the entropy gain in the solid by
\begin{eqnarray}\nonumber
    \Delta S &=& S_{\rm sol}(\rho_2,T_2) - S_{\rm sol}(\rho_1,T_1)\\\nonumber
    &=& S_{\rm liq}(\rho_1,T_1)-    S_{\rm sol}(\rho_1,T_1)\\\label{eq:dS}
    &\approx& 0.9\,k_B/\text{atom}
\end{eqnarray}

The entropy of the solid phase at a given density and temperature can be obtained from the ion-thermal contribution of the free energy in Eq.~\eqref{eq:freeenergy}, which takes the form $F_i(V,T)=3 k_BT\,\ln[\theta_0/T]$ for temperatures higher than the characteristic Debye temperature~\cite{Benedict2009}. Its derivative respect to temperature leads to
\begin{equation}\label{eq:entropy_of_solid}
    S_{\rm sol}(V,T) = 3k_B  \ln{\left(\frac{T}{\theta_0(V)}\right)} + 3 k_B,
\end{equation}
where $\theta_0$ is logarithmic moment of the phonon density of states (PDOS) at the volume $V$, defined by~\cite{wallace1972thermodynamics,Benedict2009,Robert2010},
\begin{equation}\label{eq:log_phonon_moment}
    \ln{[k_B\theta_0(V)]} = \hbar\int_0^\infty g(\omega)\,\ln{\omega}\,d\omega.
\end{equation}
Here, $g(\omega$) is the phonon density of states, and the logarithmic phonon moment, $\theta_0$, is a good approximation of the Debye temperature, $\theta(V)$.

The entropy derived from this free energy accounts only for the vibrational entropy, which dominates over the electronic entropy even at the high temperatures we are interested in. To confirm this, we calculate the electronic entropy using Mermin functional~\cite{mermin1965thermal},
\begin{equation}\label{eq:mermin}
    S_{\rm el}(T) = -k_B\int n(\epsilon)[f_i\ln{f_i}+(1-f_i)\ln{(1-f_i)}]\,d\epsilon
\end{equation}
where $n(\epsilon)$ corresponds to electronic density of states and $f_i(\epsilon)$ is the Fermi-Dirac distribution function at temperature $T$. Our calculations indicate that, at these conditions, the electronic entropy only accounts for less than 2\% of entropy of entire system.

If the melting curve is not known, one can obtain an approximate value for the melting temperature from the Lindemann criterion, which relates the melting temperature to the Gr\"uneisen parameter, $\gamma$, of the solid phase through the expression~\cite{gilvarry1956lindemann,anderson2000calculated}
\begin{equation}\label{eq:Lindemann}
    \frac{d\ln{T_m}}{d\ln{V}} = -2\left(\gamma(V) - \frac{1}{3} \right),
\end{equation}
where
\begin{equation}\label{eq:gamma_logO}
\gamma \equiv -\frac{d\ln{\theta_0(V)}}{d\ln{V}}
\end{equation}
and $\theta_0(V)$ is the logarithmic phonon moment of order $n=0$ at the volume $V$. A good approximation for $\gamma$ is to assume that it depends linearly on the volume of the solid phase, namely, $\gamma=A V+B$, which allows to obtain an analytical expression for the logarithmic phonon moment~\cite{rudindensity,Benedict2009} from Eq.~\eqref{eq:gamma_logO},
%\begin{equation}\label{eq:Gruneisen linear volume}
%    \gamma = AV + B
%\end{equation}
%Therefore we have
\begin{equation}\label{eq:thetaoverV}
    \theta_0(V) = \theta_0(V^*)\left(\frac{V}{V^*}\right)^{-B}\exp{[-A(V-V^*)]}.
\end{equation}
In the same way, it allows us to obtain an analytical expression for the melting curve from Eq.~\eqref{eq:Lindemann},
\begin{equation}\label{eq:Tmrhom}
    T_m(V) = T^* \left(\frac{V}{V^*}\right)^{-2B+\frac23} e^{-2A\left(V-V^*\right)}.
\end{equation}

We performed DFPT phonon calculations~\cite{baroni2001phonons} to obtain the PDOS, which we integrate using Eq.~\eqref{eq:log_phonon_moment} to derive the logarithmic phonon moment of the bcc phase, $\theta_0(V)$, for a number of volumes. The resulting values were used to fit the parameters $A$, $B$, $V^*$ and $\theta_0(V^*)$ in Eq.~\eqref{eq:thetaoverV}, obtaining $V^* = 6.868$ {\AA}$^3$, ${\theta}_0(V^*)$ = 1039.86~K, $A = 0.101$ {\AA}$^{-3}$, and $B = 0.515$, consistent with the values obtained by Benedict \emph{et al}~\cite{Benedict2009}. The value of $T^*$ is obtained from the melting temperature of Be at ambient conditions ($V_0=8.09$~\AA$^3$/atom) by setting $T_m(V_0) = T_0 = 1564~K$. We found $T^* = 2490~K$ a good fitting parameter.

\begin{figure}[hbt]
\centering
\includegraphics[width=8cm]{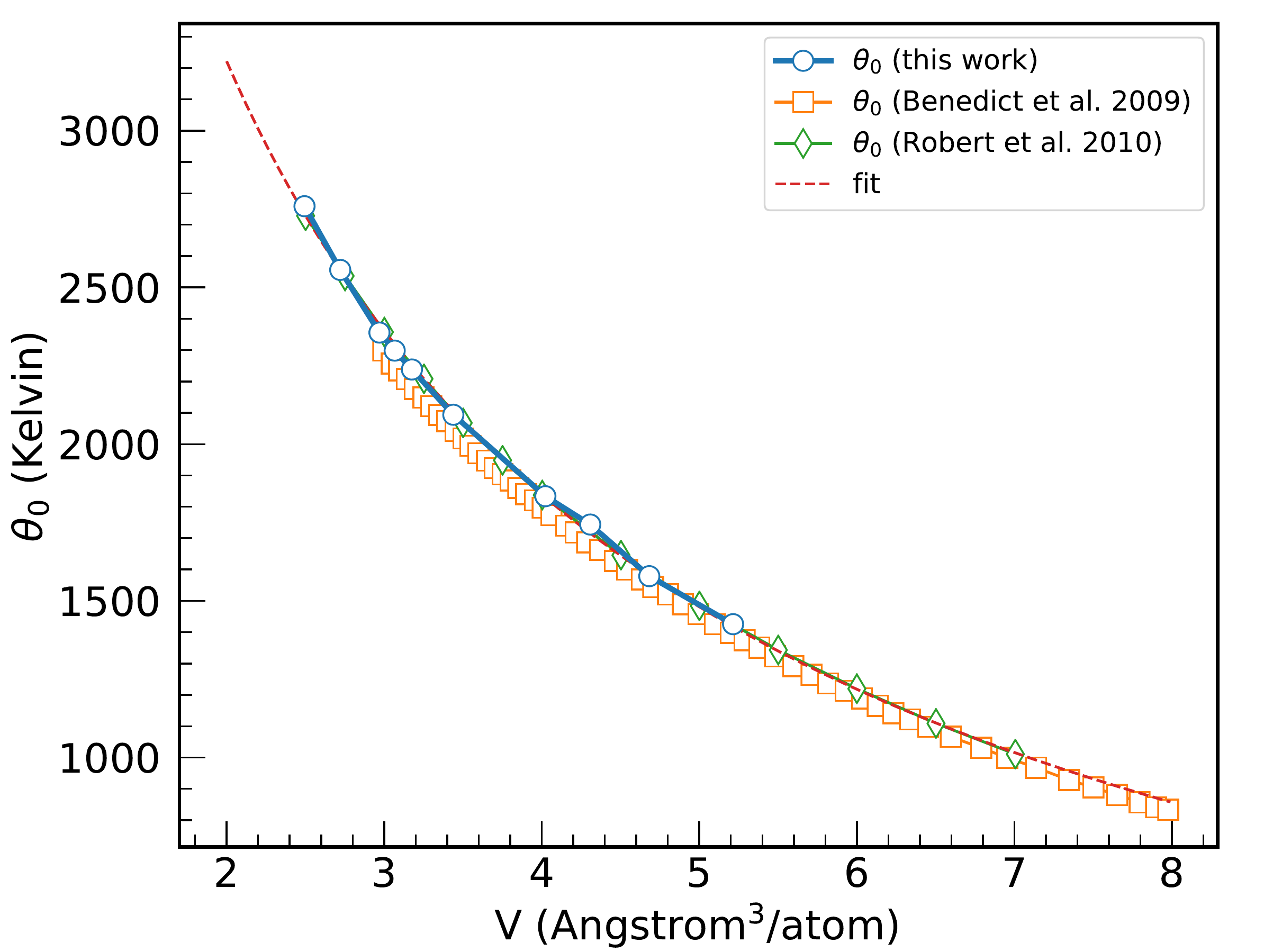}
\caption{Logarithmic phonon moments with respect of volume. Blue circle: $\theta_0$ (this work); Green diamond: $\theta_0$ by Robert \emph{et al}~\cite{Robert2010}; Orange squares: $\theta_0$ by Benedict \emph{et al}~\cite{Benedict2009}; Red dashed line: curve fit assuming Gr\"uneisen parameter $\gamma=-\frac{d\ln\theta_0}{d\ln V}$ linear with volume.}
\label{fig:logphononmoment}
\end{figure}

As shown in Fig.~\ref{fig:logphononmoment}, our values of $\theta_0$ are in good agreement with previous studies~\cite{Benedict2009,Robert2010}. The resulting melting curve obtained from Eq.~\eqref{eq:Tmrhom} is shown as the purple curve in Fig.~\ref{fig:isentrope} and it is consistent with our melting temperatures derived with TDI, which demonstrates that the approximations considered here work very well for predicting the melting temperatures.
%\begin{equation}\label{eq:Tmrhom}
%    T_m(V) = T_0 \left(\frac{V}{V_0}\right)^{-2\left(B-\frac{1}{3}\right)} e^{-2A\left(V-V_0\right)}
%\end{equation}
%Combining Eqs.~\eqref{eq:entropy_of_solid}, \eqref{eq:thetaoverV}, \eqref{eq:Tmrhom}, we obtain an approximate
We can insert Eq.~\eqref{eq:entropy_of_solid} in Eq.~\eqref{eq:dS} to relate the two melting temperatures, $T_1$ and $T_2$, with the corresponding volumes of the solid, which results in
\begin{eqnarray}\label{eq:Isentropic Gap}\nonumber
 \Delta S &=& S_{\rm sol}(V_2,T_2) - S_{\rm sol}(V_1,T_1)\\
 &=&  3\,k_B  \ln{\left(\frac{T_2}{T_1}\frac{\theta_0(V_1)}{\theta_0(V_2)}\right)}\\\nonumber
 &=& 3 k_B \ln\left[\left(\frac{V_2}{V_1}\right)^{-B+\frac23}e^{-A(V_2-V_1)}\right].
%&=& 3k_B\left(\ln(\frac{T_2}{T_2}) - \ln[\frac{\theta_0(V_2)}{\theta_0(V_1)}]\right)\\
%&=& 3k_B\left[- A(V_2-V_1) - \left(B-\frac{2}{3}\right)\ln{\frac{V_2}{V_1}}\right] 
\end{eqnarray}
This implies that $T_2=\frac{\theta_0(V_2)}{\theta_0(V_1)}e^{\Delta S/3\,k_B}T_1 $. Here $T_2\equiv T_m(V_2)$ and $T_1\equiv T_m(V_1)$ can be evaluated from Eq.~\eqref{eq:Tmrhom}, while $\theta_0(V_2)$ and $\theta_0(V_1)$ are given by Eq.~\eqref{eq:thetaoverV}. This results in a temperature difference $\Delta T=T_2-T_1$ along the isentrope given by
%\begin{eqnarray}\nonumber
%\Delta T &=& T_2-T_1\\
%&=& T_2\left[\left(\frac{V_1}{V_2}\right)^B  e^{-A(V_2-V_1)+\Delta S/3 k_B} -1\right] \\
%&=& T_2\left(1-\frac{ \theta_0(V_1)}{\theta_0(V_2)}e^{-\Delta S/3 k_B} \right) \\
%&=& T_2\left\{1-e^{2A(V_2-V_1)}(\frac{V_2}{V_1})^{2(B-\frac{1}{3})}\right\}.\\
%&=& T_m(V_2) - T_m(V_1)
%\end{eqnarray}
\begin{equation}\label{eq:DeltaT}
\Delta T= \left[\left(\frac{V_1}{V_2}\right)^B  e^{-A(V_2-V_1)+\Delta S/3 k_B} -1\right]T_1 .
%\Delta T = T^*e^{2AV^*}\left[\left(\frac{V_2}{V^*}\right)^{\frac{2-6B}{3}}e^{-2AV_2}-\left(\frac{V_1}{V^*}\right)^\frac{2-6B}{3}e^{-2AV_1}\right],
\end{equation}

Considering that the melting temperature $T_1=T_m(V_1)$ at which the isentrope intersects the melting line is known, we can infer the corresponding volume of the solid, $V_1$, from Eq.~\eqref{eq:Tmrhom}. Then, $V_2$ can be inferred from Eq.~\eqref{eq:Isentropic Gap}, assuming $\Delta S\approx 0.9\,k_B$. With these parameters, we can use Eq.~\eqref{eq:DeltaT} to determine the temperature gap between the solid and liquid isentrope with the same entropy. 
Here, $T_1=2525$~K and  $V_1=6.812$~\AA$^3$/atom (2.196~\gcc), which results in a temperature gap of $\Delta T=7500$~K. This is just slightly higher that the actual gap of 7000~K that we obtain from our TDI calculations, as we can see in Fig.~\ref{fig:isentrope}.

Therefore, the approximations that we have introduced here, based on phonon calculations coupled with a Lindemann form of the melting curve, work very well for predicting the temperature gap that arises when an isentrope intersects the melting line. We suggest that this approach can be used to estimate this gap for other materials and to predict the temperature interval over which ramp compression experiments follow the melting line.
%With $\rho_2$ = 7.000 \gcc(V$_2$ = 2.138 {\AA}$^3$/atom), we have $\rho_1$ = 2.196 \gcc, (V$_1$ = 6.812 {\AA}$^3$/atom) by formula ~\eqref{eq:Isentropic Gap} above, corresponding to 2525~K with respect to our melting line. Thus we obtained a "isentropic gap" of around 7500~K, a little bit larger than 7000~K suggested by our TDI simulations, but of similar magnitude. We could attribute this to the effect of anharmonicity at high temperature and hcp-bcc solid-solid phase transition.  Overall this approximation serves as a good reference for calculation of "isentropic gap"  and we prove that with quasi-isentropic ramp compression, we would be able to trace a section of melting line as long as 7000~K. 

%\subsection{Equation of State and Bulk Modulus}

%We can easily tell bcc beryllium holder a lower PV term. Therefore at high pressure bcc phase would be favored over hcp. 

\subsection{Phonon density of state and quasi-phonon free energy}
\label{sec:pdos}
To obtain a measure of the anharmonic effects, we
compare our free energies derived from TDI with those obtained from phonon-based methods. 
Using the \verb+Dynaphopy+ software~\cite{zhang2014phonon,carreras2017dynaphopy}, we derived the contribution to the free energy of the quasi-phonon-particles from a power spectrum of the velocity autocorrelation function,
\begin{equation}\label{eq:6}
G_{ \textbf{\textit{q}}}(\omega)=\int_{-\infty}^{+\infty}{\langle V_{ \textbf{\textit{q}}}(0)V_{ \textbf{\textit{q}}}(t)\rangle e^{i\omega t}  \, dt},
\end{equation}
where  \textbf{\textit{q}} is a wave vector in Brillouin zone and $V_{ \textbf{\textit{q}}}(t)$ is the fourier transform of weighted velocity $\sqrt{M}v(t)$ along ab-initio MD trajectory at a given time $t$. The quantity in angle brackets corresponds to the velocity autocorrelation function defined as
\begin{equation}\label{eq:vacf}
\langle V_{ \textbf{\textit{q}}}(0)V_{ \textbf{\textit{q}}}(t)\rangle = \lim_{\tau \to \infty }\frac{1}{\tau}\int_0^{\tau}V_{ \textbf{\textit{q}}}(t^{\prime})V_{ \textbf{\textit{q}}}(t^{\prime}+t) \, dt^{\prime}
\end{equation}

\begin{figure}[hbt]
\centering
\includegraphics[width=8cm]{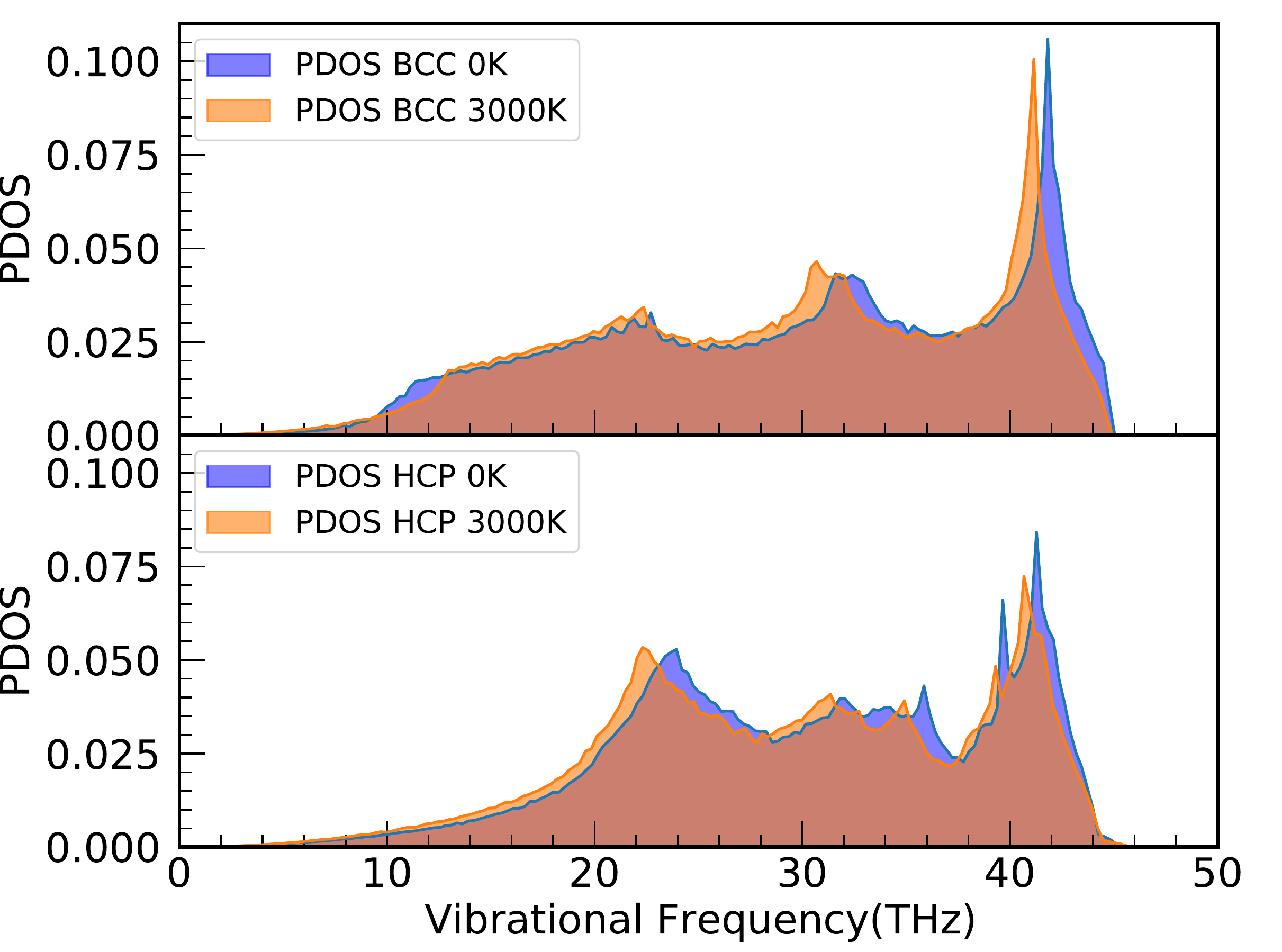}
\caption{Figure(a):Phonon vibrational density of states of both bcc Beryllium respectively at 0~K and 3000~K.Figure(b):Phonon vibrational density of states of both bcc Beryllium respectively at 0~K and 3000~K.} %{\color{red} Legend is too small / Make it 2 subplots with subplots\_adjust(hspace=0) / Use a big text for bcc and hcp, like Fig. 9 / Change yrange from 0 to 40 / set ticks at both sides / include minor ticks / Change ylabel to "Phonon density of states" }}
\label{fig:Dynaphopy}
\end{figure}

In Fig.~\ref{fig:Dynaphopy} we plot the phonon density of states (PDOS) at $T=0$~K and 3000~K from dynamic matrix. Vibrational density of state and quasi-phonon-particle fitting have been carried out with the help of phonopy~\cite{phonopy} and \verb+Dynaphopy+~\cite{carreras2017dynaphopy}. PDOS at 3000~K was derived from the quasi-phonon-particle method for both hcp and bcc phases of beryllium at $\rho=3.814$~\gcc. In contrast to the PDOS at 0~K, the PDOS of both phases at high temperature shifts towards lower frequencies, yielding phonon softening in both phases.

In order to better understand the anharmonic effects at finite temperature, we compared our entropies and free energy differences with those derived from the phonon quasiparticle method. We chose $T=3000$~K and $\rho =3.814 $ \gcc\, as reference, marked as a red cross in 
Fig.~\ref{fig:isentrope}. The entropy and free energy differences are shown in Table~\ref{table:TableI}.
\begin{table}
 \begin{tabular}{|c |c c c c|} 
%& \multicolumn{4}{c|}{Method}\\
%  \caption{Free energy difference }
 \hline
 Method              &    QHA  & Quasi-Phonon &      TDI   &\\ [0.5ex] %& TDI
 %&&phonon& &\\%  & with \\
% &&method& \\& correction
 \hline
%  $F_{\rm hcp}$ & $-$527.68 & $-$603.04      & $-$2312.94   &   $-$2317.94             \\ 
 
%  $F_{\rm bcc}$ & -538.28 & -608.58      & -2307.23   &   -2322.23            \\
 $S_{\rm hcp}\;(k_B)$ &  5.336  &   5.399      &    5.439   &\\%&       5.439    
 $S_{\rm bcc}\;(k_B)$ &  5.479  &   5.495      &    5.539    &\\%&       5.539   
 $-T\Delta S\;(meV)$ &  36.968 &   24.694      &    25.704   &\\%&       25.852    
 $F_{\rm hcp}-F_{\rm bcc}\;(meV)$ &  13.680  &   2.255      &    1.026  &\\% &       -7.579    
\hline
\end{tabular}
\caption{Comparison of the free energy difference  (in meV/atoms) between the hcp and bcc phases of Be at $T=3000$~K and $\rho=3.814$ g$\,$cm$^{-3}$, derived from different simulation methods.} 
\label{table:TableI}
\end{table}

As we can see from Table \ref{table:TableI}, at $T=$3000~K and $\rho = 3.814$~\gcc, the free energy difference between the hcp and bcc phases of beryllium given by the traditional quasiharmonic approximation method is 13.68 meV/atom. However, from the phonon quasiparticle method~\cite{sun2010lattice,zhang2014phonon,Lu2017b,Xian2019b} this difference is just 2.255~meV/atom, indicating a more stable hcp phase when anharmonic effects are taken into account, which agrees with previous experiments by either DAC~\cite{Lazicki2012b} or shock wave experiments~\cite{McCoy2019a}. From our TDI results, we obtain a free energy difference between hcp and bcc beryllium of 1.026 meV/atom, close to the result suggested by the quasi-phonon method. Therefore anharmonic effects lower the free energy of hcp structure by more than 10~meV at 3.814 \gcc and 3000~K, which helps to explain the higher hcp-bcc transition pressure in our phase diagram Fig.~\ref{fig:PhaseDiagramComplete}. Thus, the anharmonic effects captured by TDI are well approximated by the quasiphonon method.

We further investigated the entropy of both phases at these condition. Our results, summarized in Table~\ref{table:TableI}, show that the entropic term in the Gibbs free energy difference, $G_{\rm hcp} - G_{\rm bcc}$, at $T=3000$~K is 25.7~meV/atom from our TDI calculations, 11~meV/atom smaller than that derived from QHA. Thus, the anharmonic effects on the entropy are stronger in the hcp structure compared to bcc, enhancing the stability of hcp structure.

%when no k-point correction is applied, meaning that bcc structure is more stable than hcp. After the correction, this difference reduces increases to 2.53 meV/atom, meaning that hcp is in fact more stable.

\subsection{Electronic Density of States}
\begin{figure}[hbt]
\centering
\includegraphics[width=8cm]{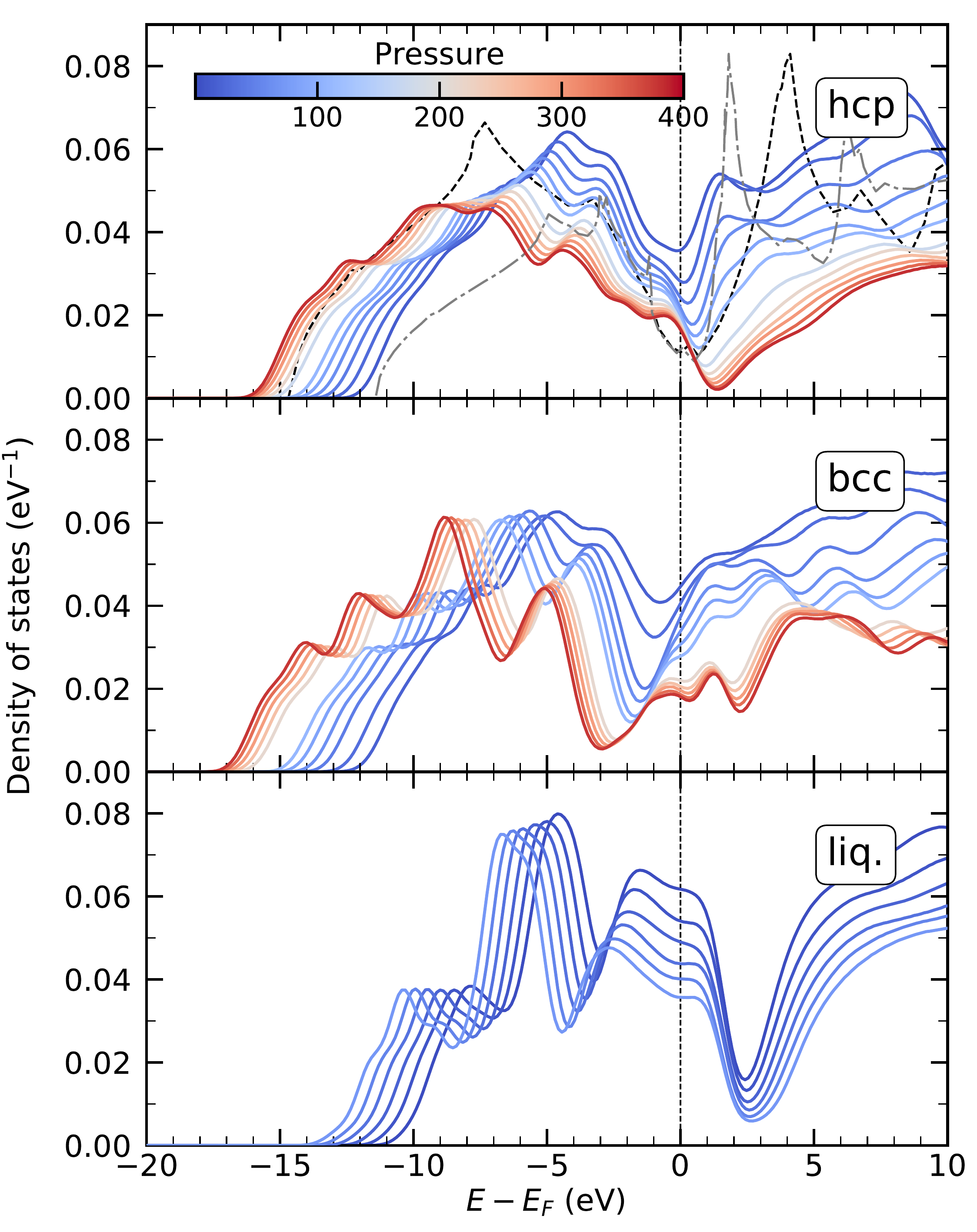}
\caption{Electronic density of states of Be at 3000 K. DOS of hcp Be at 0 and 200~GPa from Ref.~\cite{Guo2014} are shown in dot-dashed and dashed lines in the upper panel.}
\label{fig:DOSBe}
\end{figure}

%\begin{figure}[hbt]
%\centering
%\includegraphics[width=8cm]{DOS.pdf}
%\includegraphics[width=8cm]{DOS_Be_300GPa_v1.png}
%\includegraphics[width=8cm]{DOS_Be_300GPa_v2.png}
%\caption{Electronic Density of states at 300 GPa.}
%\label{fig:DOSBe2}
%\end{figure}

In Fig.~\ref{fig:DOSBe} we show how the electronic density of states (DOS) of Beryllium changes with pressure in the hcp, bcc, and liquid phases at 3000 K.
We obtained the DOS from the analysis of the eigenenergies provided by Kohn-Sham DFT, as we have done in previous works~\cite{Gonzalez-Cataldo2020,Soubiran2019}. With a Brillouin zone sampled by Monkhorst-Pack method with $2\times2\times2$ $k$-points sampling~\cite{Monkhorst1976b}, we obtained smooth DOS curves by averaging over the MD-simulation snapshots and by applying a Gaussian smearing of 0.1 eV to the band energies. The DOS at every snapshot was aligned at its respective Fermi energy, and then we averaged all of them together. The average Fermi energy was then subtracted out and the integrated DOS was normalized to 1.

The K shell (1s) electrons form a sharp peak (not displayed in the figure) centered around 100 eV below the Fermi energy. The conduction band, formed by the L shell electrons, shows similar features for both the bcc and hcp phases.
While the DOS of Be shows a minimum around the Fermi energy for the hcp phase, this minimum is shifted to energies above the Fermi energy in the liquid phase.  In our EOS, we did not find signatures of pressure ionization that can be attributed to an energy minimum~\cite{GonzalezMilitzer2020,Gonzalez2020}, and no gap-opening occurred in our electronic DOS for the regime of pressures explored. However the density of states near the Fermi energy does decrease under compression, leaving a possibility for a gap opening at higher pressures.

%As we can see from the equation of state, the pressure was almost linear with respect %to the internal energy. Thus Gruneisen parameter scales proportional to volume, %independent from temperature.
%\begin{figure}[hbt]
%\centering
%\includegraphics[width=8cm]{Gruneisen.pdf}
%\caption{Gruneisen parameter}
%\label{fig:Gruneisen}
%\end{figure}

\section{Conclusion}
\label{IV}

We performed a systematic investigation of the beryllium phase diagram under extreme conditions using the first principles thermodynamic integration method. At 0~K, we find that the hcp phase of beryllium transforms to the bcc phase at 405~GPa and that at higher temperatures, the Clapeyron slope of the hcp-bcc phase boundary is negative. We showed that the quasiharmonic approximation tends to underestimate the stability of the hcp phase. When the full anharmonicity is considered, we find a solid-solid phase boundary that is similar to that predicted by QHA at low temperatures, but is shifted to higher pressures with increasing temperature. Our triple point is located at 164.7~GPa and 4314~K, much higher pressure and temperature than the 85~GPa and 3400~K suggested by the quasiharmonic approximation.

By fitting the Fourier transform of the velocity autocorrelation function to obtain the phonon quasiparticles, we obtained the vibrational density of states at 3000~K and calculated corresponding free energy. The free energy difference between hcp and bcc phases calculated by the TDI method is much smaller than that derived by QHA and agrees well with the phonon quasiparticle method, consistent with our predictions of a larger hcp domain and higher hcp-bcc transition pressures.

We calculated the shock Hugoniot curve and found it to be consistent with previous shock experiments. Our Hugoniot crosses the solid-solid phase boundary at 200~GPa and 4000~K with small drop in temperature and encounters the melting line at 235~GPa and 4900~K. Previous shock Hugoniot experiments that predict a lower pressure for the onset melting from sound speed measurements may correspond to a case of ``cold melting,'' followed by recrystallization, yielding a premature measurement of the melting point.

Our melting line shows good agreement with two phase method and heat-until-it-melts simulations predictions below 6000~K, and is lower than that predicted by TDI simulations using the empirical MEAM model. Our derived melting temperatures extend the melting curve of beryllium up to a pressure of 1600 GPa.

%The temperature gap along the melting curve between the solid and liquid isentropes is as large as 7000~K giving a theoretical prediction of future ramp compression experiments.
%Our isentrope of both liquid and solid phase of beryl-lium  hold  a  large  gap  of   7000  kelvin  along  the  melt-ing line, giving a possible theoretical prediction of futureramp compression experiment.
We computed isentropes in the liquid and solid phases and found them to be shallower than our melting curve in pressure-temperature space. We predict the thermodynamic path of a hypothetical quasi-isentropic ramp compression experiment. Starting with a liquid isentrope, it would follow the melting line while the sample is in a solid-liquid mixed state before entering the solid phase. We predict Be to remain in a solid-liquid mixed state to be present over a large temperature interval of 7000~K. Based on the canonical value for the entropy of fusion, 0.9 $k_B$/atom, one can expect for other materials the solid-liquid state to be present over several thousand Kelvin. The magnitude also depends on the shape of the melting curve.

\begin{acknowledgments}
This work was in part supported by the National Science Foundation-Department of Energy (DOE) partnership for plasma science and engineering (grant DE-SC0016248) and the University of California Laboratory Fees Research Program (grant LFR-17-449059). FGC and BM acknowledge support from DOE-National Nuclear Security Administration (grant DE-NA0003842). Computational resources at the National Energy Research Scientific Computing Center were used. R. Jeanloz, F. Soubiran, and B.K. Godwal  provided constructive comments.
\end{acknowledgments}

%\nocite{*}

%\bibliography{Be_TDI.bib}% Produces the bibliography via BibTeX.
%\bibliographystyle{prb}

%\bibliographystyle{apsrev4-1}

\end{document}